\documentclass{aa}
\usepackage{graphicx}
\usepackage{natbib}
\newcommand{\h}{$^{\rm h}$}
\newcommand{\m}{$^{\rm m}$}
\newcommand{\s}{$^{\rm s}$}
\begin{document}
   \title{Spectral characteristics of water megamaser galaxies II}
   
   \subtitle{\object{ESO\,103--G035},
   \object{TXS\,2226--184}, and \object{IC\,1481} \thanks{Based on observations collected at the European Southern Observatory, Chile}} 

   \author{N. Bennert\inst{1}
          \and
          H. Schulz\inst{1}\fnmsep\inst{2}\thanks{deceased}
          \and
          C. Henkel\inst{3}}

   \offprints{Nicola Bennert}

   \institute{Astronomisches Institut Ruhr--Universit\"at Bochum,
              Universit\"atsstra{\ss}e 150,
              D--44780 Bochum, Germany;
             \email{nbennert@astro.rub.de}
         \and Universidad Cat\'olica del Norte, 
       Avenida Angamos 0610, Antofagasta, Chile
\and Max--Planck--Institut f\"ur Radioastronomie, Auf dem H\"ugel 69,
   D--53121 Bonn, Germany; \email{chenkel@mpifr-bonn.mpg.de}}

   \date{Received 13 October 2003 / Accepted 25 February 2004}

  \abstract{
Long--slit optical emission--line spectra of the H$_2$O megamaser galaxies  
\object{ESO\,103--G035}, \object{TXS\,2226--184}, and \object{IC\,1481} 
are evaluated in order to look for characteristics
typical for water--megamaser galaxies. We present rotation curves, line ratios, 
electron densities, temperatures, and H$\beta$ luminosities. The successful 
line--profile decompositions rest on $d$--Lorentzians with an additional parameter 
$d$ to adjust the wings, rather than Gaussians or Lorentzians as basic
functions. No significant velocity gradient is found along the major axis in the
innermost 2\,kpc of \object{TXS\,2226--184}. 
\object{IC\,1481} reveals a spectrum suggestive of a vigorous starburst in the 
central kiloparsec 10$^8$ years ago. None of the three galaxies shows any
hints for outflows nor 
special features which could give clues to the presence of H$_2$O megamaser emission. 
The galaxies are of normal Seyfert--2 (\object{ESO\,103--G035}) or LINER 
(\object{TXS\,2226--184}, \object{IC\,1481}) type.
\keywords{Galaxies: active --
          Galaxies: nuclei --
          Galaxies: individual: \object{ESO\,103--G035}, \object{TXS\,2226--184}, \object{IC\,1481}}}

\titlerunning{Spectral characteristics of three water megamaser galaxies}

\authorrunning{Bennert et al.}

   \maketitle

\section{Introduction}
Extragalactic H$_2$O megamaser sources are orders of magnitude 
more powerful than strong $\lambda$=1.3\,cm Galactic masers of the same type and have been 
discovered in about three dozen galactic nuclei, almost all of them of Seyfert--2 
or LINER (low--ionization nuclear emission--line region) 
type (e.g. \citet{bra96}; \citet{gre03}).

According to the so--called unified model \citep{ant85,ant93}, Seyfert--2 objects 
contain a nuclear molecular torus seen edge--on that surrounds an X--ray luminous, 
winds and jets producing active galactic nucleus (AGN). It is plausible that this
geometry is favorable for megamaser activity, because the necessary large column 
densities of warm ($T \ge 400$\,K) and dense ($n$(H$_2$) $\ge 10^{7}$\,cm$^{-3}$) molecular gas 
enriched with H$_2$O molecules are likely to be supplied. Interaction of molecular 
gas with a radio jet may also cause conditions favorable for the occurrence of H$_2$O 
masers \citep{pec03}.

This paper is the second one in a series analyzing
the spectral properties of water--vapor megamaser 
galaxies at optical wavelengths. 
In the first one \citep{sch03}, emission--line profiles of IC\,2560, 
NGC\,1386, NGC\,1052, and Mrk\,1210 were evaluated. Galactic rotation and outflow of 
narrow--line gas are common features of this sample of water--megamaser galaxies.
All decomposed line--systems exhibit AGN typical line ratios. For NGC\,1052 and 
Mrk\,1210, maser emission triggered by the optically detected outflow components cannot
be ruled out.

In this work, we continue to explore the structure of galactic nuclei containing H$_2$O 
megamaser sources (\object{ESO\,103--G035}, \object{TXS\,2226--184}, \object{IC\,1481}) in order to look for 
distinguishing characteristics and to obtain clues to the nuclear geometry.
\subsection{Individual objects}
\citet{mar79} and \citet{pic82} classified {\bf \object{ESO\,103--G035}} (IRAS\,18333--6528) as 
a Seyfert--1.9 galaxy whereas \citet{mor88} did not find any hints of broad line wings.
Observations with EXOSAT (European Space Agency's X-ray Observatory) revealed variable, 
strong absorption of soft X--rays \citep{war88}. \citet{hei94} find \object{ESO\,103--G035} 
among the warmest far--infrared emitters in their sample of IRAS (Infrared Astronomical 
Satellite) galaxies with spectral energy distributions peaking near 60 microns. It 
unveals a high excitation emission--line spectrum superposed on a red stellar continuum 
with absorption lines. \citet{bra96} discovered H$_2$O megamaser emission in this galaxy.

So far, the most luminous known H$_2$O maser is found in the galaxy {\bf \object{TXS\,2226--184}} 
(IRAS\,F22265--1826; \citet{koe95}). Hubble Space Telescope (HST) and Very Large Array 
observations from \citet{fal00} suggest that it is a galaxy of type later than S0 with an 
inclination of $\sim$\,70\degr. Their H$\alpha$ + [\ion{N}{ii}] map exhibits a gaseous, 
jetlike structure perpendicular to the galaxy's major axis and its nuclear dust lane.
The 8.4\,GHz radio continuum map 
shows emission that is elongated in the same direction 
as the H$\alpha$ + [\ion{N}{ii}] emission. \citet{fal00} concluded that the nuclear accretion 
disk, the obscuring torus, and the large--scale molecular gas layer are roughly coplanar.
\citet{tay02} found \ion{H}{i} in absorption towards \object{TXS\,2226--184}, consisting of two 
components with widths of 125\,km\,s$^{-1}$ and 420\,km\,s$^{-1}$. They suggest that the 
\ion{H}{i} absorption is produced in the central parsecs of the galaxy, on a scale similar 
to that which gives rise to the water maser emission. 

Little is known about {\bf \object{IC\,1481}} (IRAS\,23168+0537) which was classified as LINER by 
\citet{huc82}. \citet{bra96} discovered its water megamaser emission. Due to its amorphous 
appearance (e.g. \citet{fal01}) 
its inclination is hard to determine: \citet{vau91} (RC3) list an inclination of 
$i$ $\sim$\,70\degr~whereas $i$ $\sim$\,30\degr~is given in \citet{bra97}. \citet{van02} 
classify \object{IC\,1481} in their catalogue of host galaxies of supernovae as a peculiar Sb
galaxy. 
\section{Observations and data reduction}
The spectra described here were obtained by H. Domg{\"o}rgen in May 1996 using the Boller 
\& Chivens spectrograph attached to the Cassegrain focus of the European Southern Observatory 
(ESO) 1.52\,m telescope. Observations were made in two spectral ranges (3400--5400\,\AA~and 
5400--7400\,\AA) through the nucleus along the major axis of each galaxy with exposure times 
ranging between 1800\,s and 3600\,s. A log of the observations is given in Table~\ref{log}.
The detector used was La Silla CCD No. 39 (Loral Lesser) with 15\,$\mu$m wide square pixels.
The spatial resolution element is 0\farcs68\,pix$^{-1}$. Seeing and telescope properties 
limited the spatial resolution to the range 1\farcs2 $-$ 2\farcs5 which was determined by 
the width of spectra of standard stars on the focal exposures. The 2\arcsec~wide slit projects 
to a spectral resolution of $\sim$\,2.7\,\AA~($\sim$\,130\,km\,s$^{-1}$) as is confirmed by the
full width at half--maximum (FWHM) of comparison lines and the [\ion{O}{i}]\,$\lambda$6300 
night--sky line.

Standard reduction including bias subtraction and flat--field correction was 
performed using the ESO MIDAS\footnote{Munich Image Data Analysis System, trade--mark of the ESO} 
software (version Nov. 98). Night--sky spectra ``above'' and ``below'' any notable galaxy 
emission were interpolated in the region of the galactic spectrum and subtracted in each 
case. Excellent flux interpolation was achieved by rebinning the spectra to a scale of 
0.97\,\AA~per pixel during wavelength calibration. The curve of \citet{tug77} was used to 
correct for atmospheric extinction. The spectra were flux calibrated using the standard 
stars LTT\,7987 or CD\,--32\degr 9927.

From the frames cleaned in this way, single rows were extracted (each row corresponding to 
0\farcs68 and 2\arcsec~along and perpendicular to the slit direction, respectively). Along 
the ``spatial axis'' of the CCD, we identified the ``photometric center'' (that we choose 
as the ``zero'' on the spatial scale) with the most luminous row of the CCD (``central row'');
this does not have to coincide with the dynamical center. NGC\,3115 or NGC\,4179 (both 
classified as S0 galaxies in the NASA/IPAC Extragalactic Database (NED)) were used as templates 
to subtract the stellar contribution from each single row. These template spectra turned out 
to be rather similar in our spectral ranges and were usually suitable to fit the absorption 
lines of the megamaser galaxies. Only for \object{IC\,1481}, a special treatment was necessary that 
will be discussed in the next section.

For line ratios presented here, the three central rows were averaged to enhance the S/N without
loosing any spatial information. Hence, the line ratios refer to the central
region of $2\arcsec \times 
2\arcsec$.  Forbidden--line wavelengths were taken from \citet{bow60}. Heliocentric corrections 
as given in Table~\ref{log} were added to the observed velocities. The sample properties are 
given in Table~\ref{sample}.
\begin{table*}
\begin{minipage}{180mm}
\caption[]{\label{log}Summary of the spectroscopic observations}
\begin{flushleft}
\begin{tabular}{lccccccc}
\hline
\hline
\rm{Object/hel. corr.\footnote{This heliocentric correction was added to the
measured radial velocities.}} & \rm{Date (beg.)} &
\rm{p.a. (slit)} & $\lambda$
\rm{range} (\AA) &
\rm{Exp. time} (s)\\
\hline
\rm{ESO\,103-G035} & 14-{\rm May}-96 & 44\degr & 5400-7400 & 1800\\
$+14$\,km/s  & 17-{\rm May}-96 & 44\degr & 3400-5400 & 1800\\
\hline
\rm{TXS\,2226-184} & 14-{\rm May}-96 & 50\fdg3 & 5400-7400 & 2700\\
$+29$\,km/s & 17-{\rm May}-96 & 50\fdg3 & 3400-5400 & 2700\\
\hline
\rm{\object{IC\,1481}} & 14-{\rm May}-96 & 24\degr & 5400-7400 & 3600\\
$+25$\,km/s &  17-{\rm May}-96 & 24\degr & 3400-5400 & 2400\\
\hline
\end{tabular}
\end{flushleft}
\end{minipage}
\end{table*}
\begin{table*}
\begin{minipage}{180mm}
\caption[]{\label{sample}H$_2$O--megamaser sample properties}
\begin{flushleft}
\begin{tabular}{lccc}
\hline
\hline
 & \rm{ESO\,103-G035} & \rm{TXS\,2226-184} & \rm{\object{IC\,1481}}\\
\hline
& \multicolumn{3}{c}{\rm literature}\\
\hline
{\rm coordinates\footnote{
J2000, NED}} & 18\h 38\m 20\s~-65\degr 25\arcmin 39\arcsec 
& 22\h 29\m 13\s~-18\degr 10\arcmin 47\arcsec & 23\h 19\m 25\s~+05\degr 54\arcmin 22\arcsec\\
\rm{incl.\footnote{Inclination angle from face--on (RC3)}(\degr)} & 67\degr 
& 70\degr & 62\degr\\
\rm{m.a.\footnote{Position angle of major axis (RC3)}}(\degr) & 44\degr & 51\degr & 70\degr
(30\degr\footnote{Taken from \citet{bra97}})\\
$v_{\rm
opt}$\footnote{Heliocentric velocity from RC3} (\rm{km} \rm{s}$^{-1}$)
& 3983 $\pm$ 34
& -- & 6118\footnote{Taken from \citet{van02}} $\pm$ 50 \\
\rm{Type\footnote{Classification and Hubble type (NED)}} & \rm{Sy1.9/2, Sa0} &
\rm{?, S0?} & \rm{LINER, 
Sb pec$^{\rm f}$}\\
\hline
& \multicolumn{3}{c}{\rm this work}\\
\hline
$v_{\rm hel}$\footnote{Heliocentric velocity as derived from our rotation curves.} 
(\rm{km} \rm{s}$^{-1}$) & 4040 $\pm$ 20 & 7520 $\pm$ 30 & 6120 $\pm$ 30\\
\rm{dist.\footnote{Distance via direct application
of $v = H_0 \cdot d$ ($H_0$ = 65\,km\,s$^{-1}$\,Mpc$^{-1}$) using velocities
relative to the 3K background as derived from $v_{\rm hel}$ with the velocity calculator
provided by NED.}} (\rm{Mpc})
& 61.9 & 110.6 & 88.5 \\
\rm{lin. sc.} (\rm{pc/\arcsec}) & 300.2 & 536.3 & 428.8\\
$M_{\alpha}$\footnote{Estimated lower limit of mass
with $\alpha$=0.6 \citep{leq83}
within the central 2.8\,kpc (diameter).} (10$^9$\,M$_{\sun}$) 
& 1.2 & -- & --\\
\rm{Type} & Sy2 & LINER & LINER\\
\hline
\end{tabular}
\end{flushleft}
\end{minipage}
\end{table*}
\section{Results}
\begin{figure}[h!]
\centering
\includegraphics[width=6cm,angle=-90]{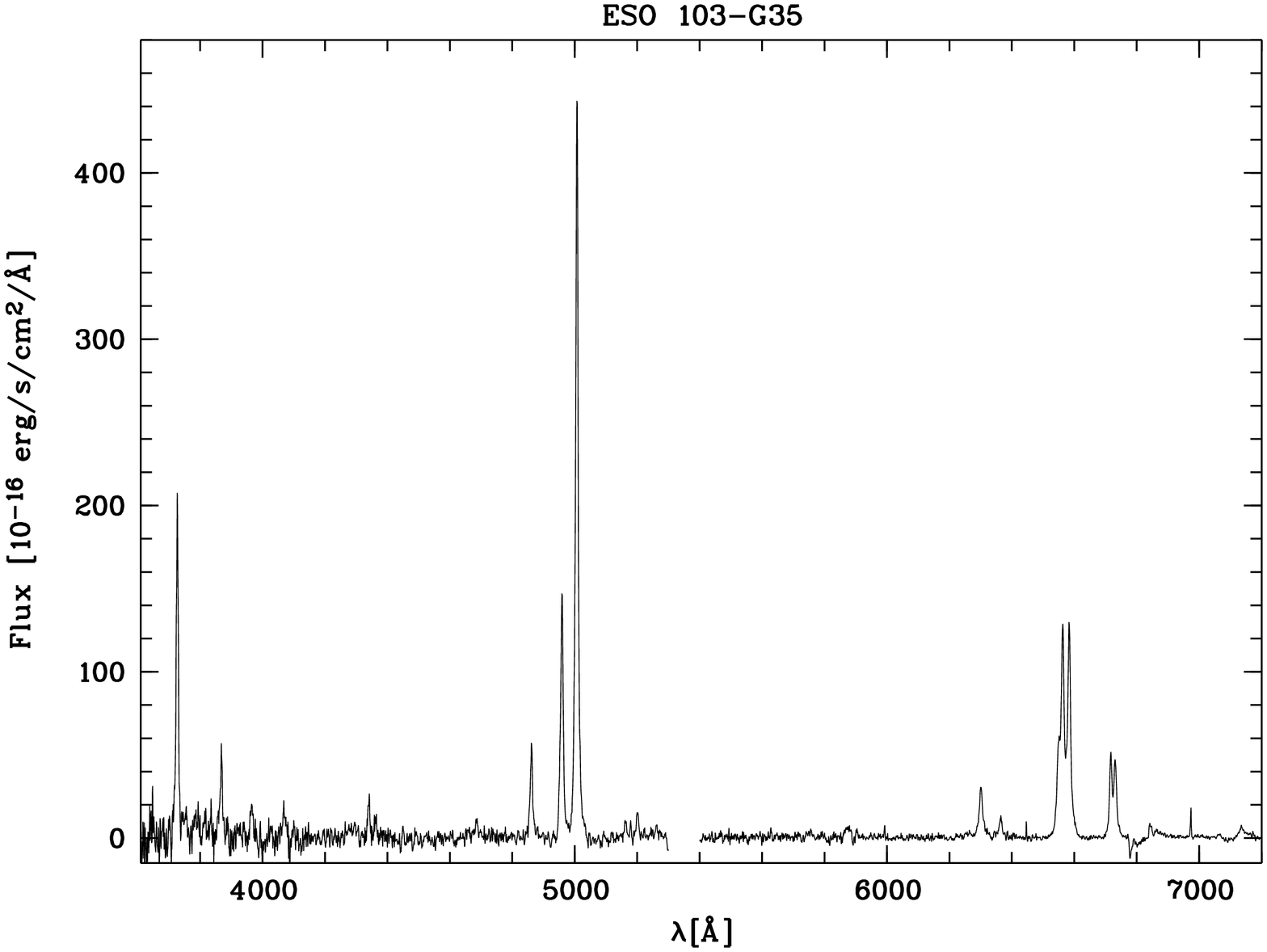}
\includegraphics[width=6cm,angle=-90]{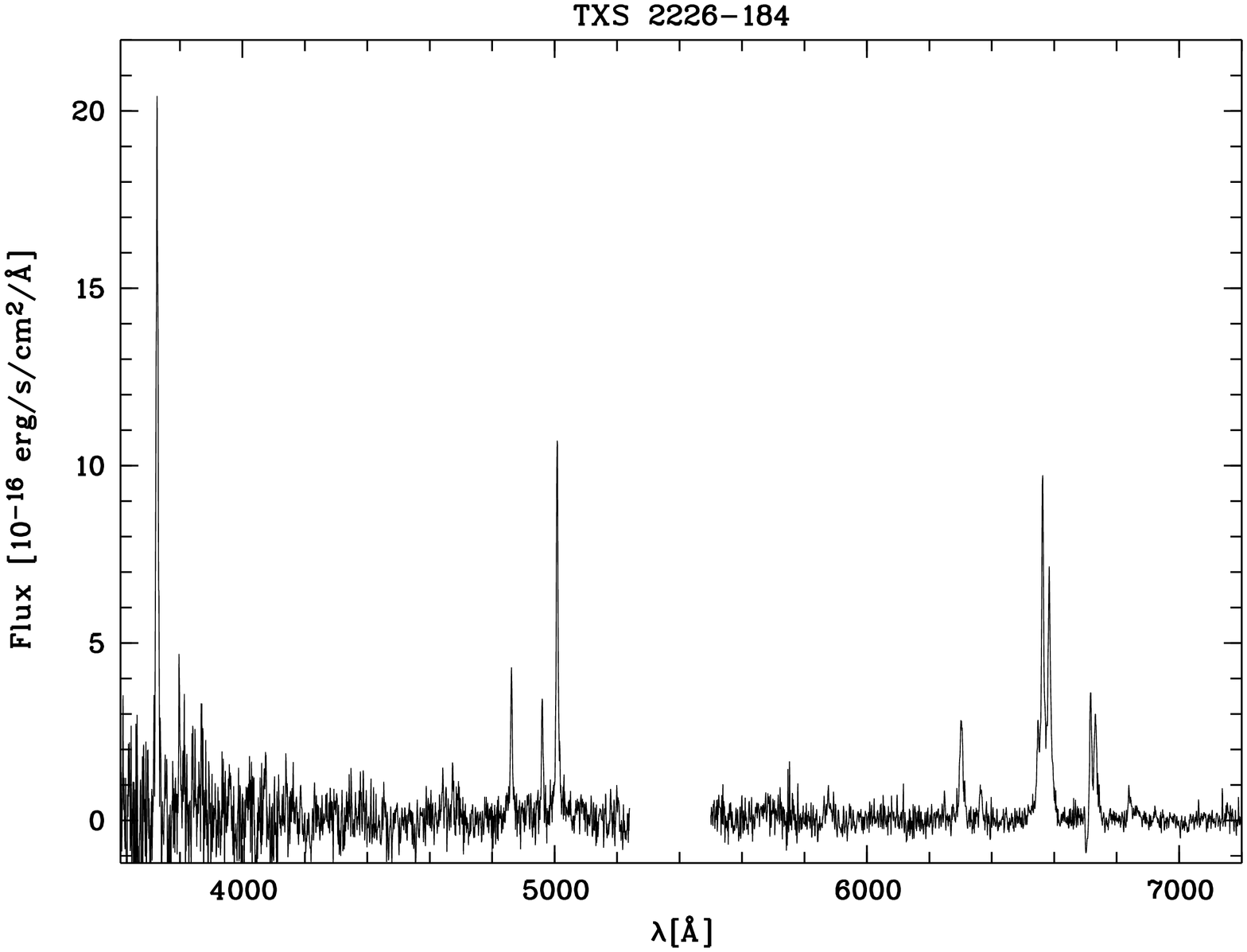}
\includegraphics[width=6cm,angle=-90]{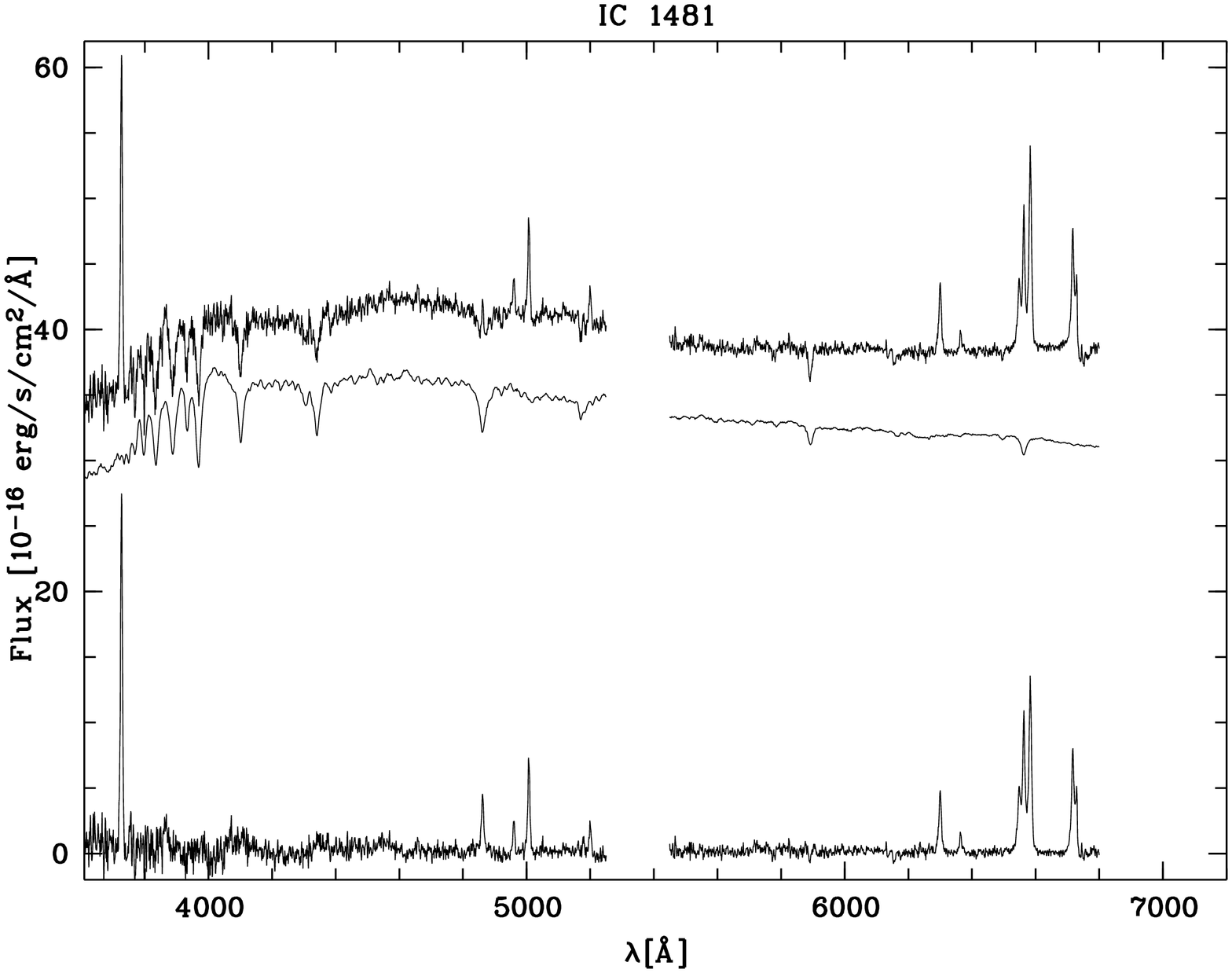}
\caption[]{\label{specblue}Combined dereddened spectra of the red and blue wavelength 
range rebinned to rest wavelengths for comparison. A template galaxy spectrum 
(NGC\,3115 or NGC\,4179) was 
subtracted for \object{ESO\,103--G035} and \object{TXS\,2226--184}. The upper observed spectrum of \object{IC\,1481} 
shows strong Balmer absorption 
lines and weak Ca\,II H+K features. Thus, a linear combination of 
A0V $-$ F8V stars and the template galaxy  was subtracted (see middle spectrum). The result
is shown in the lower spectrum. (Both upper spectra of \object{IC\,1481} are shifted
vertically by an arbitrary amount.)}
\end{figure}

The dereddened spectra of the innermost regions ($2\arcsec \times 2\arcsec$)
rebinned to rest wavelengths are shown in Fig.~\ref{specblue}.
\subsection{Stellar population of the galactic nuclei}
To derive emission--line fluxes, we need to model the 
stellar population which can contaminate our emission--line spectra
due to underlying absorption lines. 
As a first approximation,
we simply subtracted one of the S0--type template spectra scaled to the 
inter--emission--line parts from all spectra.
For two of the three objects, 
there remains no notable continuum nor absorption--line spectrum. 
This means that, to our accuracy, the stellar populations in the nuclei of \object{ESO\,103--G035} 
and \object{TXS\,2226--184} closely resemble the old stellar populations in NGC\,3115 and 
NGC\,4179.

\citet{jog01} 
classified \object{ESO\,103--G035} as ``Seyfert 2b''
meaning that they observed high order Balmer lines in
absorption which are often interpreted to
indicate a significant young stellar population.
We do not find high order Balmer absorption lines in our spectra.
This might be due to our lower S/N spectra: \citet{jog01} used the same
telescope and instrument but a UV--blazed grating and
a CCD camera with high quantum efficiency below
5000\,\AA. Furthermore, they used twice the integration
time with half the spectral resolution.
However, having a closer look at their online data,
we find that \object{ESO\,103--G035} resembles
more closely some galaxies which \citet{jog01} classified as ``Seyfert 2a''
(spectra showing both emission and absorption lines
in which the Balmer series is {\it not} detected in absorption, e.g. ESO\,323+G32)
than ``Seyfert 2b'', thus in agreement with our spectra.
In our raw data we see \ion{Ca}{ii} K\,$\lambda$3933, weak CN\,$\lambda$4200,
G band\,$\lambda$4301 and \ion{Mg}{i}\,$\lambda$5175 absorption lines.
Since all of them can be well fit by NGC\,3115, we exclude 
the possibility of a significant residual starburst contribution.

For \object{IC\,1481}, the subtraction of S0--type spectra
was not satisfying as strong Balmer absorption lines and weak \ion{Ca}{ii} H+K features 
remained. 
Thus, we modelled the underlying stellar continuum in two ways:
i.) by using the P\'EGASE code (version 2.0)
by \citet{fio97} and
ii.) by using the best fit of linear combinations of main sequence A to F 
stars and the template spectrum of NGC\,3115.
The latter provided a much better fit to the observed spectra
of \object{IC\,1481}, and was thus used. The fit was performed as follows:

We tried a variety of linear combinations of main sequence A to F 
stars on the one hand (taken from the catalogue of \citet{pic98}, rebinned with a spline 
function to the higher resolution of our spectra), and the template spectrum of 
NGC\,3115 on the other hand to fit the blue spectrum with its strong features and 
simultaneously the spectral distribution in the red range. The scaled monochromatic 
fluxes at 5556\,\AA~were weighted according to the different fluxes of A0V $-$ F8V stars,
and, by using a Salpeter initial mass function, normalized to the flux and mass of an 
A5V star.

After a first approximation of such a linear combination to gain the absorption--line 
free H$\alpha$/H$\beta$ value, the spectra of \object{IC\,1481} were dereddened using the 
recombination value for the intensity ratio H$\alpha$/H$\beta$ = 3.1 (a typical value 
for AGNs) and an average reddening curve (\citet{ost89}, Table 7.2; MIDAS command 
``extin/long''). This procedure was carried out several times to find the best combination 
iteratively ($E_{\rm B-V}$ = 0.15):
\begin{eqnarray}
\label{com}
f_{\rm \object{IC\,1481}} 
= 0.099\,f_{\rm A0V} + 0.076\,f_{\rm A2V} 
+ 0.076\,f_{\rm A3V}\nonumber\\
+ 0.071\,f_{\rm A5V}   
+ 0.064\,f_{\rm A7V} + 0.061\,f_{\rm F0V} \nonumber\\
+ 0.039\,f_{\rm F5V} + 0.036\,f_{\rm F8V}
+ 0.48\,f_{\rm NGC\,3115}\nonumber\\
= 8 \cdot 10^{-16}\,\rm{erg}\,\rm{s}^{-1}\,\rm{cm}^{-2}\,\AA^{-1} \hspace*{0.2cm} .
\end{eqnarray}
As the lowest spectrum in Fig.~\ref{specblue} shows, there are no significant 
absorption features left that could constrain a more elaborate population model to be 
subtracted. A slight depression near $\sim 4020$\,\AA~might be suggestive of 
\ion{He}{I}\,$\lambda$4026 from early B stars but is too noisy for detailed fitting. There is 
no evidence for \ion{He}{I}\,$\lambda$4471 so that we estimate a main--sequence turnoff at late 
B stars. This yields a rough age of an assumed instantaneous starburst of $\sim 1.7 
\cdot 10^8$ years \citep{ibe67}.  

We converted the theoretical physical continuum fluxes of the above stars as tabulated
by \citet{cox00} into fluxes which would be measured at the assumed distance of $r$ = 
88.5\,Mpc (\object{IC\,1481}, Table~\ref{sample}) using $f_{\lambda} = F_{\lambda} \frac{R^2}{r^2}$
($R = 1.2$ $-$ $2.9$\,R$_{\sun}$) and compared them with the measured ones. This leads to 
9.4 $\cdot 10^7$ stars or a mass of $1.7 \cdot 10^8 $\,M$_{\sun}$ in the range A0V to F8V. 
In a more complete range for the original starburst (O5V $-$ M5V, mass range 
60 $-$ 0.21\,M$_{\sun}$), this would correspond to a total mass of newborn 
stars of $9.9 \cdot 10^8$\,M$_{\sun}$ or 
1.4 $\cdot 10^9$ stars. Although these estimates are rather crude, they 
nevertheless show that there was a vigorous starburst in the central kiloparsec 
of \object{IC\,1481} $10^8$ years ago. This finding is contradictory to \citet{boi00}
who claimed that in their sample of central regions of 12 galaxies (normal galaxies, 
starburst galaxies, LINERs, Seyfert 1s \& 2s), LINERs show the oldest population with 
little star formation still going on.
It is, however, not inconsistent with an
HST image that was analyzed by \citet{fal01}. They 
speculated that the irregular shape of the galaxy might be the site of an
ongoing galaxy merger.

It could be interesting to perform a detailed empirical population synthesis
as proposed and developed by \citet{bic88}, \citet{sch96}, \citet{fer01},
and \citet{fer03} for our three
galaxies and to compare the results with other stellar populations of
Seyfert--2 and LINER--host galaxies (e.g. \citet{sch99,boi00,jog01}).
However, on the one hand, having a very small sample (one Sy2 and two LINERs), 
we could not provide deeper insights
into the links between nuclear activity and
the star formation history of the host galaxy.
On the other hand, as we
do not expect to have missed a significant contribution of
underlying stellar absorption lines, this time--consuming
procedure is beyond the scope of this work.
\subsection{Rotation curves}
\begin{figure}[h!]
\centering
\includegraphics[width=6.cm,angle=-90]{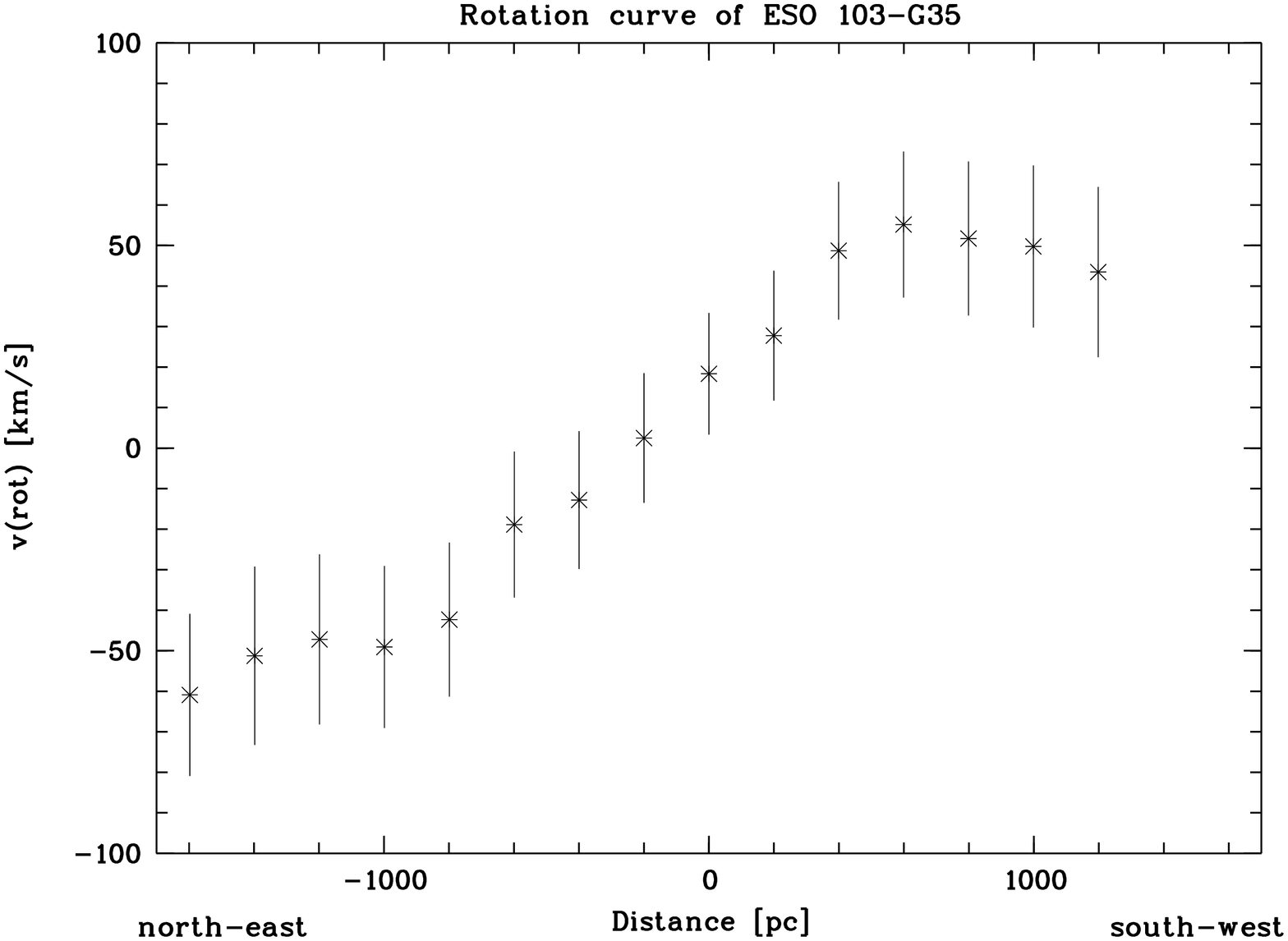}
\includegraphics[width=6.cm,angle=-90]{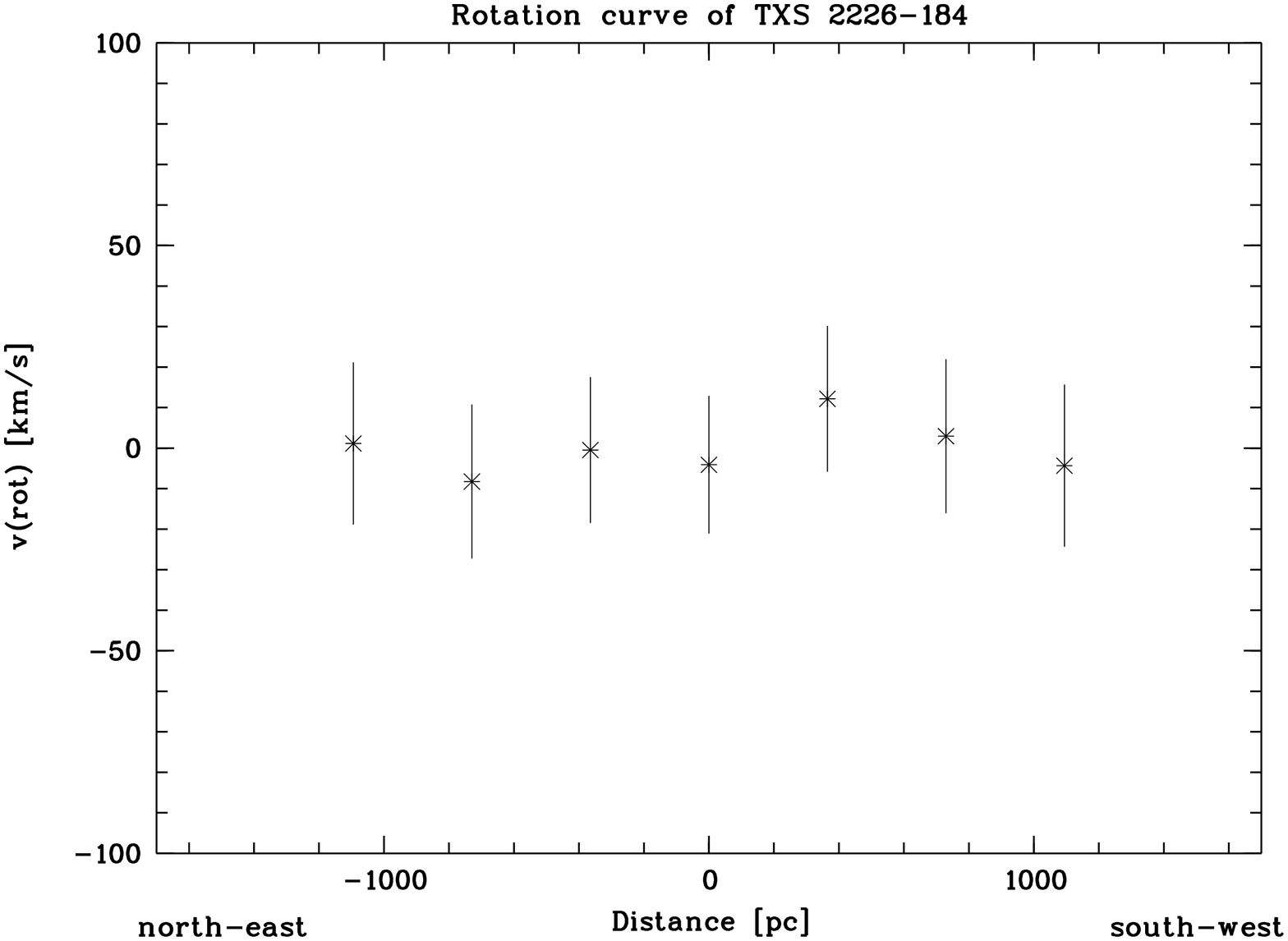}
\includegraphics[width=6.cm,angle=-90]{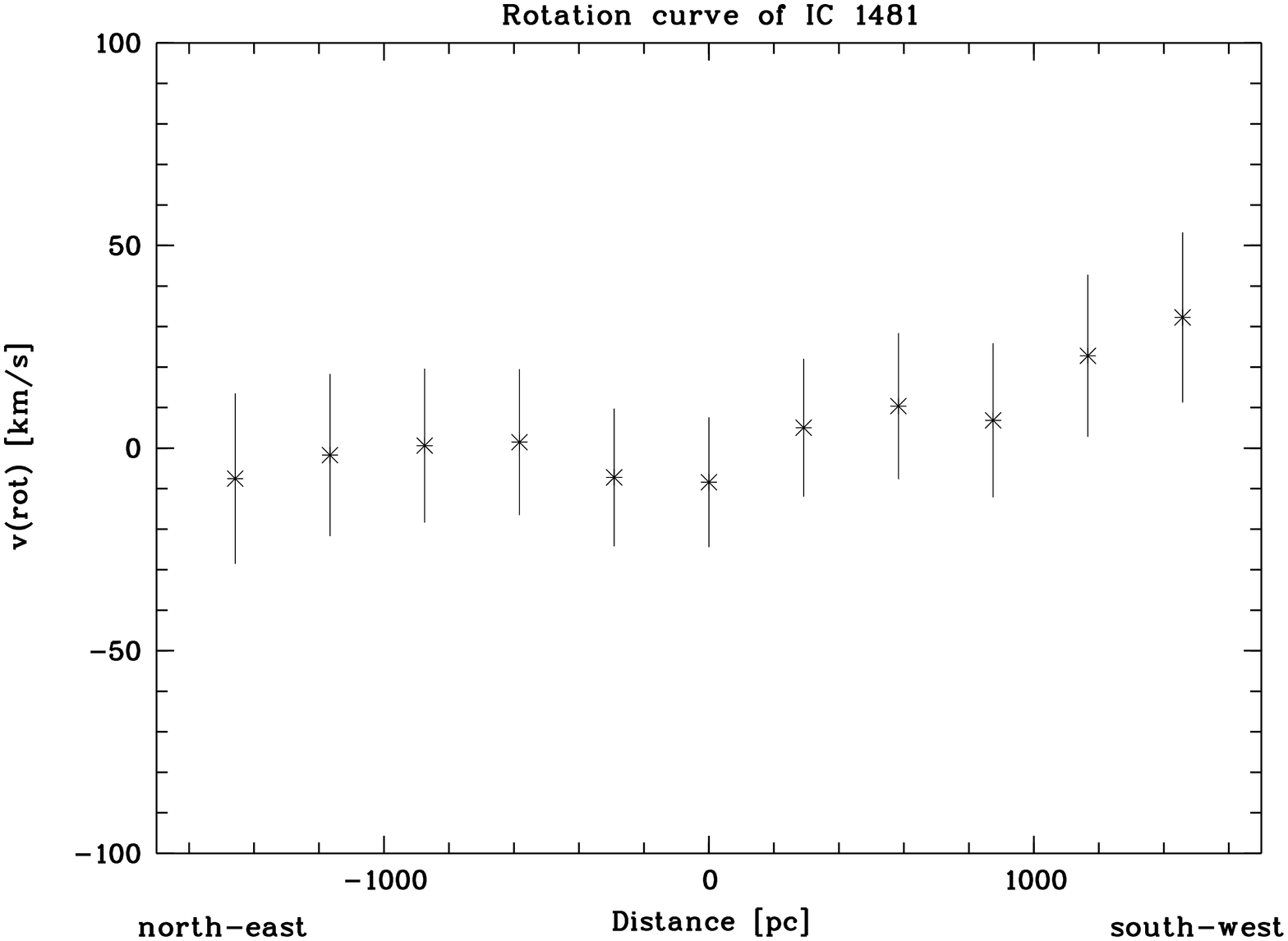}
\caption[]{\label{velocity}
Rotation curves of the
three H$_2$O--megamaser galaxies derived from the average value
of the H$\alpha$ and [\ion{N}{ii}]\,$\lambda$6583 peak wavelengths.}
\end{figure}

To gain velocity curves, we took averaged peak wavelengths of H$\alpha$ and 
[\ion{N}{ii}]\,$\lambda$6583 (with heliocentric correction applied, see Table~\ref{log}) 
and subtracted the heliocentric systemic velocity derived by symmetrising the curve of each 
galaxy. This heliocentric velocity can be found in Table~\ref{sample}. Rotational velocities 
were calculated assuming that all spectra were taken along the major axis of each galaxy. 
Thus, the observed velocities had to be corrected simply 
for the inclination angles (as taken from 
RC3): 
\begin{eqnarray}
v_{\rm rot} = \frac{v_{\rm obs}}{\sin i} \hspace*{0.2cm} .
\end{eqnarray}
For the central 2 $-$ 4\,kpc for which the H$\alpha$ and [\ion{N}{ii}]\,$\lambda$6583 lines
were strong enough to be measured, the rotational velocities are presented in 
Fig.~\ref{velocity}.
The central mass could only be estimated for \object{ESO\,103--G035} (see Table~\ref{sample}), because
the ``rotation'' curves for \object{IC\,1481} and \object{TXS\,2226--184} are remarkably flat. As a lower limit, 
we estimate the mass as
\begin{eqnarray}
M = \alpha \cdot \frac{v_{\rm rot}^2 \cdot R}{G}
\end{eqnarray}
with $\alpha$ = 0.6, taking into account non--Keplerian motion at those locations, where the 
gravitational potential consists of the superposition of a flat disk and a spherical component 
\citep{leq83}.
The resulting mass of \object{ESO\,103--G035} is 1.2 $\times$ 10$^{9}$\,M$_{\sun}$ within the central
2.8\,kpc (Table~\ref{sample}).

The flat ``rotation'' curve of \object{IC\,1481} can be explained by its amorphous
appearance: The spectra 
were possibly not observed along the major axis, since the position of the major axis is very 
uncertain. The observations were done at a position angle of $\sim$\,24$\degr$
corresponding approximately to the position angle reported by
\citet{bra97}. However, in the RC3,
the position angle of the major axis is listed as 70$\degr$
(Table~\ref{sample}).

\object{TXS\,2226--184} was definetely observed along the major axis. This galaxy might 
therefore be an elliptical with its central gas disk lying coplanar to the major axis. While 
the latter is strenghtened by observations of the narrow--line region, the surface brightness
distribution favors a spiral galaxy \citep{fal00}. The flat rotation curve may thus be caused 
by the fact that our observations are confined to the extent of the nuclear bulge where gas 
kinematics do not follow the trend expected for a highly inclined disk that should be dominant 
further out.
\begin{table*}
\begin{minipage}{180mm}
\caption[]{\label{fwhm}
Heliocentric velocities and line widths (FWHM) (separated by $|$) in
km\,s$^{-1}$ as well as mean $d$--parameter as
derived from $d$--Lorentz fits for the central row.}
\begin{flushleft}
\begin{tabular}{lccc}
\hline
\hline
line & \rm{ESO\,103-G035} & \rm{TXS\,2226-184} & \rm{\object{IC\,1481}}\\
\hline
H$\beta$                      & 4011 $|$ 479 & 7498 $|$ 446 & 6138 $|$ 445\\
$[\ion{O}{iii}]\,\lambda$5007 & 4045 $|$ 539 & 7553 $|$ 397 & 6123 $|$ 405\\
$[\ion{O}{i}]\,\lambda$6300   & 4063 $|$ 584 & 7564 $|$ 676 & 6087 $|$ 405\\
H$\alpha$                     & 4046 $|$ 504 & 7486 $|$ 327 & 6098 $|$ 392\\
$[\ion{N}{ii}]\,\lambda$6583  & 4046 $|$ 443 & 7486 $|$ 526 & 6097 $|$ 416\\
$[\ion{S}{ii}]\,\lambda$6731  & 4059 $|$ 411 & 7515 $|$ 389 & 6107 $|$
363\footnote{$[\ion{S}{ii}]\,\lambda$6731 was truncated by telluric absorption
  bands and thus, $[\ion{S}{ii}]\,\lambda$6716
  was used instead.}\\
\hline
mean    & 4045$\pm$8$ | $493$\pm$26 & 7517$\pm$14 $|$ 460$\pm$51 &
6108$\pm$7 $|$ 404$\pm$11\\
mean $d$ & 1.6$\pm$0.2 & 1.1$\pm$0.1 & 1.4$\pm$0.3\\
\hline
\end{tabular}
\end{flushleft}
\end{minipage}
\end{table*}
\begin{table*}
\begin{minipage}{180mm}
\caption[]{\label{ratio}
Observed and reddening--corrected 
line intensity ratios relative
to H$\beta$.\footnote{Derived
from the central region ($2\arcsec \times 2\arcsec$)}}
\begin{flushleft}
\begin{tabular}{lcccccc}
\hline
\hline
Line & \multicolumn{2}{c}{\rm ESO\,103-G035} 
& \multicolumn{2}{c}{\rm TXS\,2226-184} & \multicolumn{2}{c}{\rm \object{IC\,1481}}\\
& \multicolumn{1}{c}{$F_{\rm obs}$} & \multicolumn{1}{c}{$F_{\rm dered}$} 
& \multicolumn{1}{c}{$F_{\rm obs}$} & \multicolumn{1}{c}{$F_{\rm dered}$}
& \multicolumn{1}{c}{$F_{\rm obs}$} & \multicolumn{1}{c}{$F_{\rm dered}$}\\
\hline
$[\ion{O}{ii}]\,\lambda3727$ & 1.53 & 2.56 & 4.39 & 5.69 & 3.03 & 3.43\\
$[\ion{Ne}{iii}]\,\lambda3868$ & 0.67 & 1.03 & 0.4 & 0.49 & 0.51 & 0.57\\
H$\epsilon+[\ion{Ne}{iii}]\,\lambda$3967 & 0.22 & 0.33 & -- & -- & -- & -- \\
$[\ion{O}{iii}]\,\lambda$4363 & 0.12 & 0.15 & -- & -- & -- & --\\
$\ion{He}{ii}\,\lambda$4686 & 0.14 & 0.15 & 0.18 & 0.19 & -- & --\\
$[\ion{O}{iii}]\,\lambda$5007 & 9.93 & 9.44 & 2.81 & 2.74 & 1.04 & 1.02\\
$[\ion{Fe}{ii}]+[\ion{Fe}{vii}]\,\lambda$5159 & 0.13 & 0.12 & -- & -- & -- & --\\
$[\ion{Fe}{vi}]\,\lambda$5176 & 0.05 &  0.04 & 0.14 & 0.13 & 0.13 & 0.12\\
$[\ion{N}{i}]\,\lambda$5199  & 0.18 & 0.15 & 0.3 & 0.28 & 0.31 & 0.3\\
$\ion{He}{i}\,\lambda$5875 & 0.28 & 0.18 & 0.58 & 0.47 & -- & -- \\
$[\ion{O}{i}]\,\lambda$6300 & 1.62 & 0.93 & 2 & 1.42 & 0.96 & 0.83\\
H$\alpha$ & 6.11 & 3.1 & 4.45 & 3.1 & 3.66 & 3.1\\
$[\ion{N}{ii}]\,\lambda$6583 & 6.4 & 3.25 & 5.13 & 3.57 & 2.36 & 2\\
$[\ion{S}{ii}]\,\lambda$6716 & 2.14 & 1.05 & 1.56 & 1.05 & 1.5 & 1.26\\
$[\ion{S}{ii}]\,\lambda$6731 & 2.1 &  1.03 & 1.48 & 0.98 & -- & -- \\
$\ion{He}{i}\,\lambda$7065 & 0.06 &  0.03 & -- & -- & -- & --\\
$[\ion{Ar}{iii}]\,\lambda$7136 & 0.49 &  0.22 & -- & -- & -- & --\\
\hline
\end{tabular}
\end{flushleft}
\end{minipage}
\end{table*}
\begin{table*}
\begin{minipage}{180mm}
\caption[]{\label{result}
Reddening--corrected H$\beta$ luminosity and results from dereddened line ratios.}
\begin{flushleft}
\begin{tabular}{lccc}
\hline
\hline
 & \rm{ESO\,103-G035} & \rm{TXS\,2226-184} & \rm{\object{IC\,1481}}\\
\hline
$L_{\rm H\beta}$\footnote{Derived
from the central region ($2\arcsec \times 2\arcsec$)} (10$^{40}$\,erg\,s$^{-1}$) & 2.65
& 0.49 & 0.59\\
$A_{\rm V}$$^a$ & 1.9 & 1. & 0.47 \\
$n_{\rm e}$$^a$ (cm$^{-3}$) & 600 $\pm$ 50 & 550 $\pm$ 100 & --\\
$T$\footnote{Derived from 
the central row ($0\farcs68 \times 2\arcsec$)} (K) & 13000 $\pm$ 500 & -- & --\\
$[\ion{O}{iii}]\,\lambda5007/[\ion{O}{ii}]\,\lambda3727$ & 
3.69 & 0.48 & 0.3\\
$[\ion{O}{iii}]\,\lambda5007/[\ion{O}{i}]\,\lambda6300$ & 
10.15 & 2.08 & 1.23\\
$[\ion{O}{i}]\,\lambda6300/$H$\alpha$ & 
0.3 & 0.46 & 0.27\\
$[\ion{S}{ii}]\,\lambda6716+\lambda6731/$H$\alpha$ & 
0.67 & 0.66 & 0.79\footnote{Using an estimated ratio of
$[\ion{S}{ii}]\,\lambda6716/\lambda6731$ $\sim$\,1.07 for a typical 
$n_{\rm e}$ of $\sim$\,550\,cm$^{-3}$.}\\
$[\ion{N}{ii}]\,\lambda6583/$H$\alpha$ & 
1.05 & 1.15 & 0.65\\
\hline
\end{tabular}
\end{flushleft}
\end{minipage}
\end{table*}

\subsection{Emission--line profiles, intensities and widths}
Attempts to fit the emission--line profiles of the strongest lines from H, [OIII], 
[OI], [NII], and [SII] in the template subtracted spectra were made with Gauss 
functions, Lorentz functions (= Cauchy distributions), and modified Lorentzians,
dubbed here as $d$--Lorentzians, 
as described in Appendix~\ref{app}.
(Note, that in the MIDAS environment, Lorentz functions 
are called Cauchy functions, while $d$--Lorentzians are called Lorentz functions.) 
 
\begin{figure}[t!]
\centering
\includegraphics[width=6cm,angle=-90]{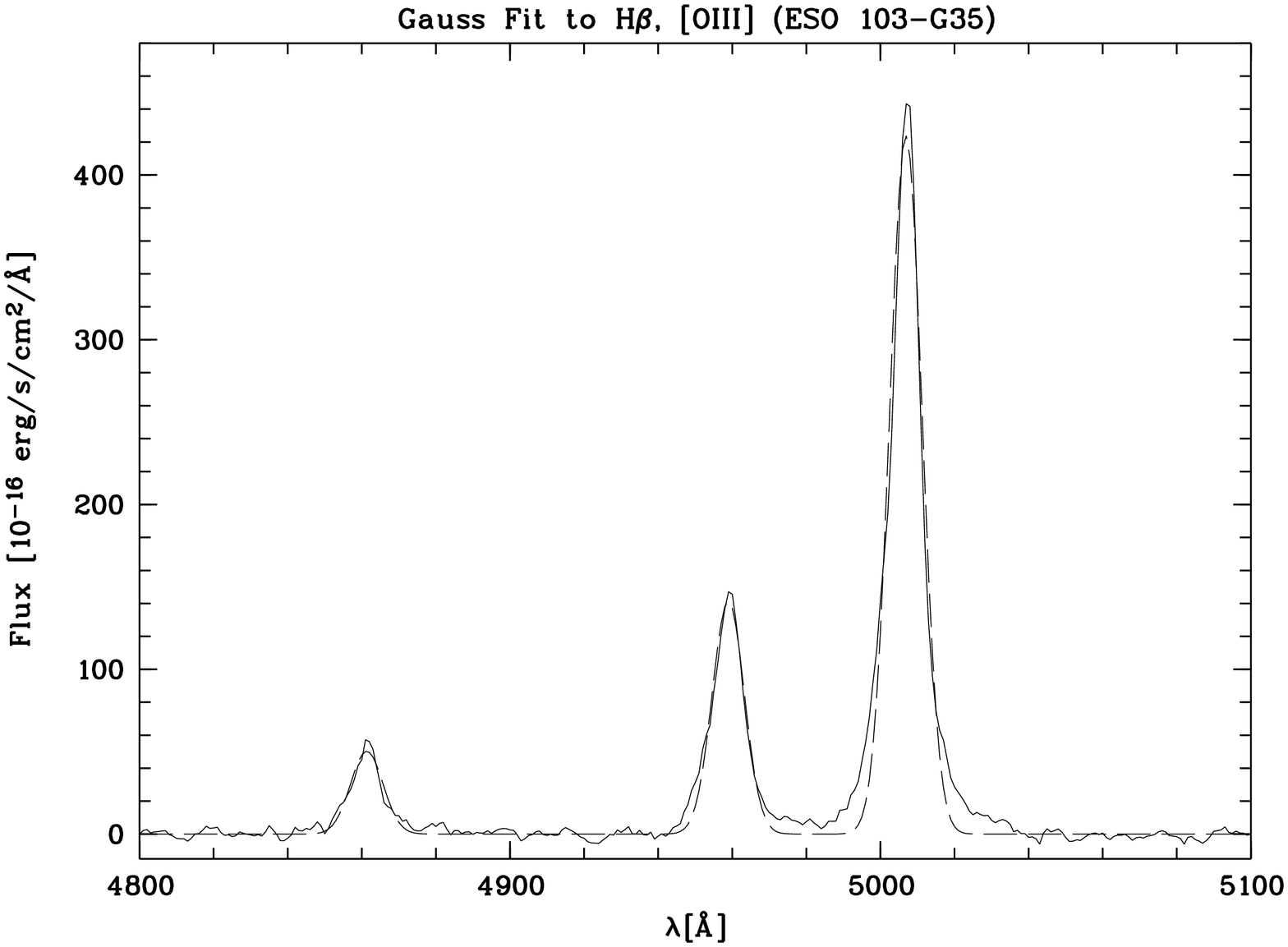}
\includegraphics[width=6cm,angle=-90]{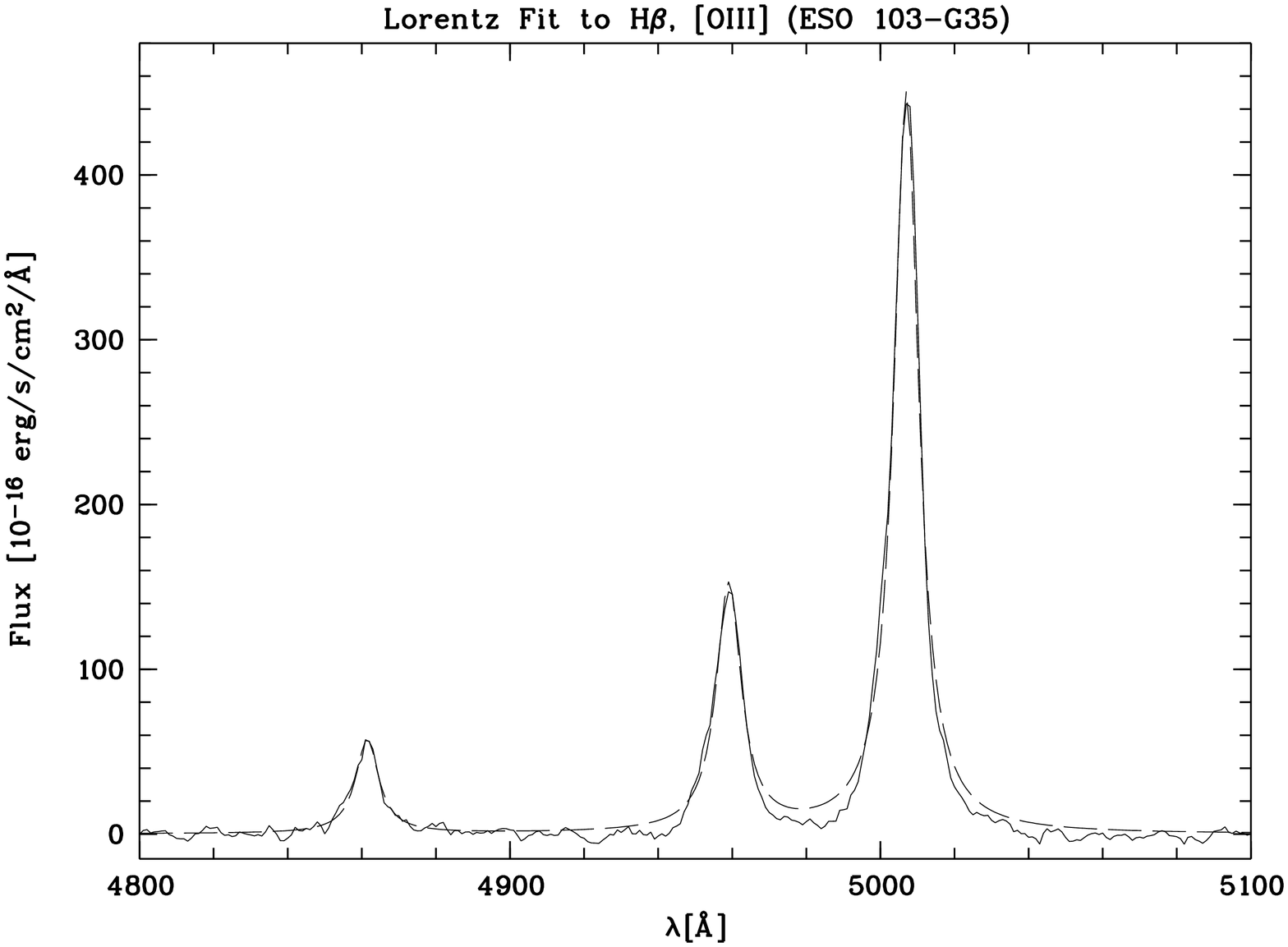}
\includegraphics[width=6cm,angle=-90]{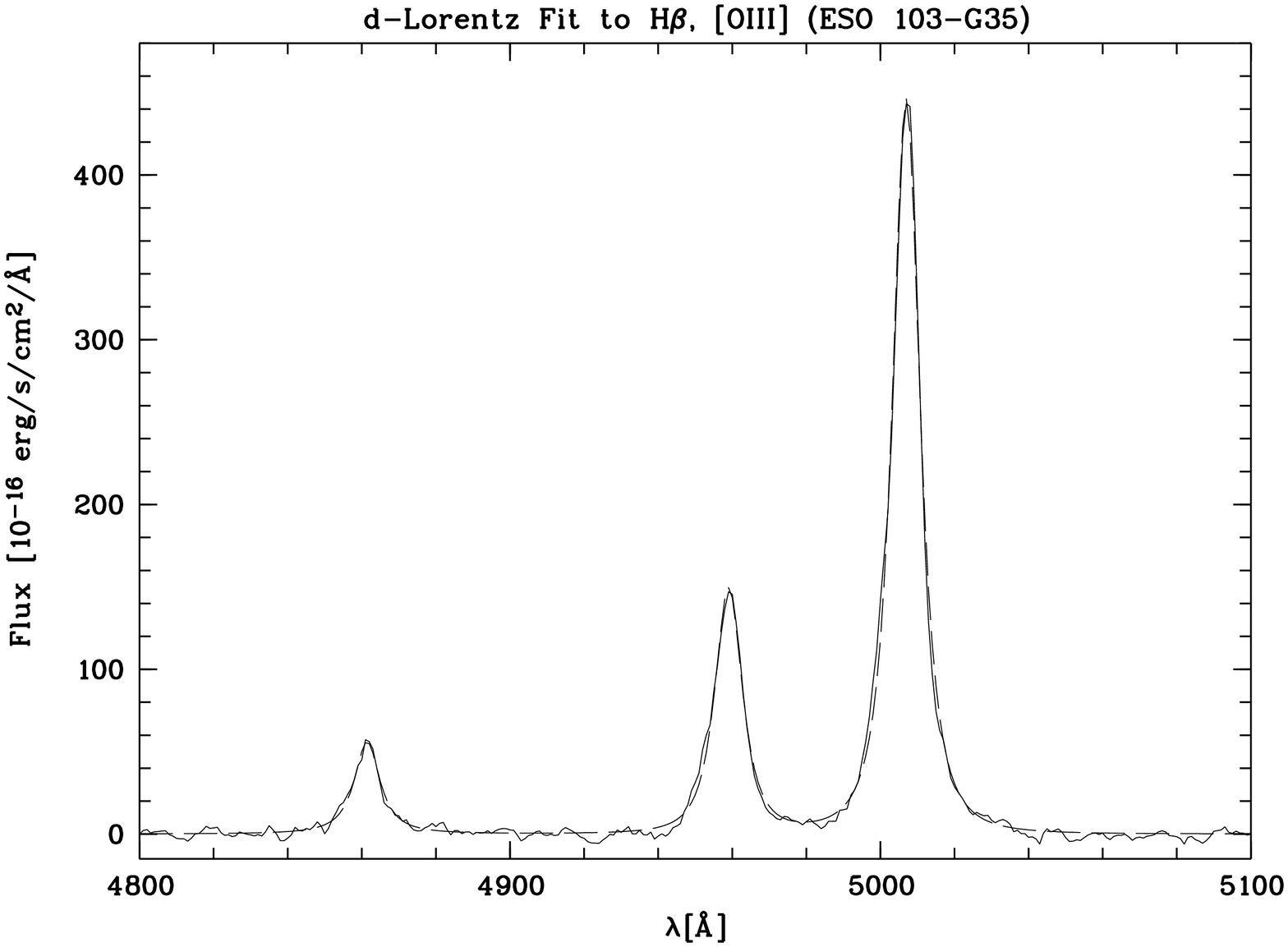}
\caption{\label{hb_o3}
H$\beta$ and 
[\ion{O}{iii}]\,$\lambda\lambda$4959,5007 
fit by three Gaussians (upper), three Lorentzians (middle) and
three $d$--Lorentzians (lower)
(\object{ESO\,103--G035}).
The total fit is shown (dashed line).
Gaussians yield too narrow wings, Lorentzians
too broad ones, whereas
$d$--Lorentzians give the best total fit.}
\end{figure}

\begin{figure}[h!]
\centering
\includegraphics[width=6cm,angle=-90]{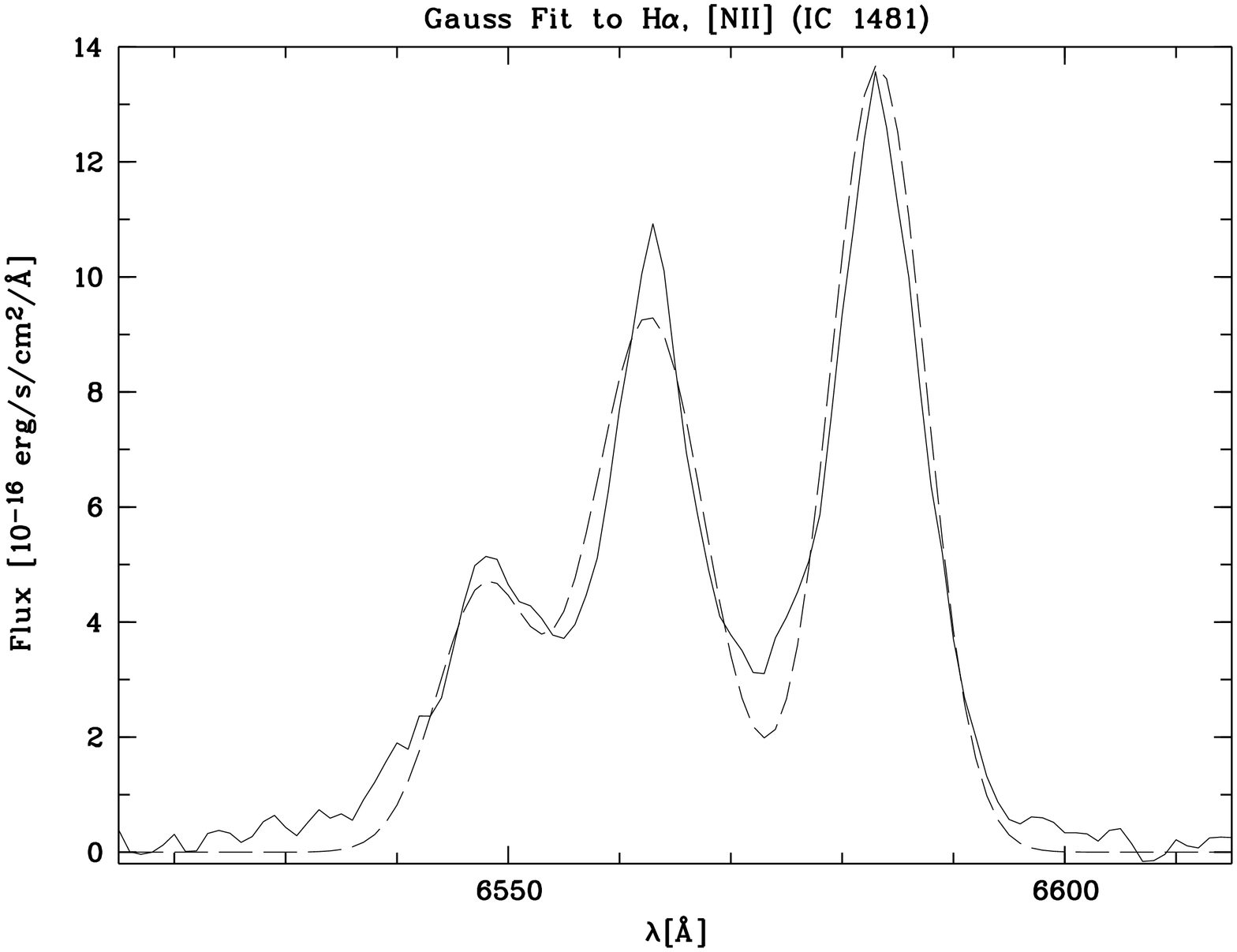}
\includegraphics[width=6cm,angle=-90]{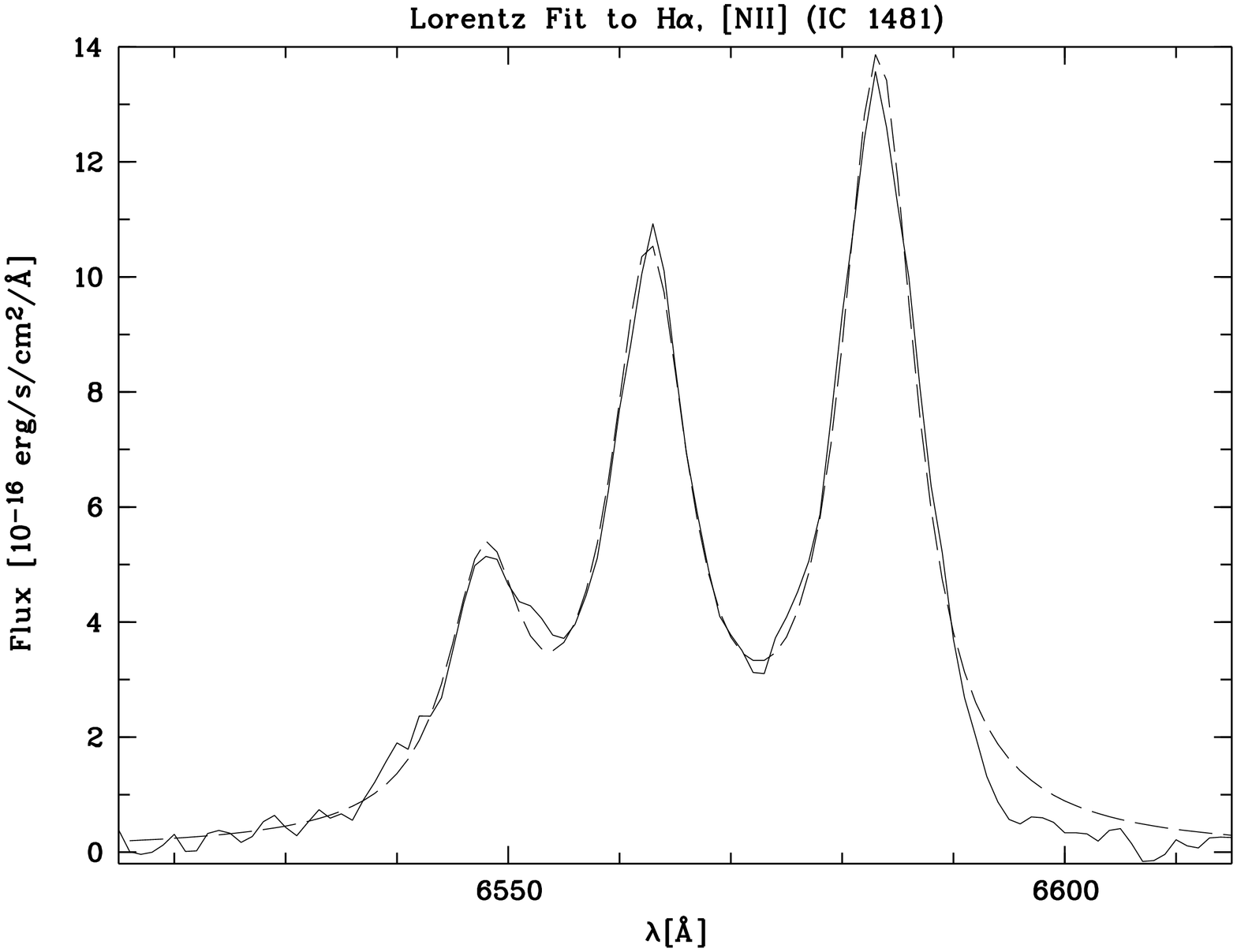}
\includegraphics[width=6cm,angle=-90]{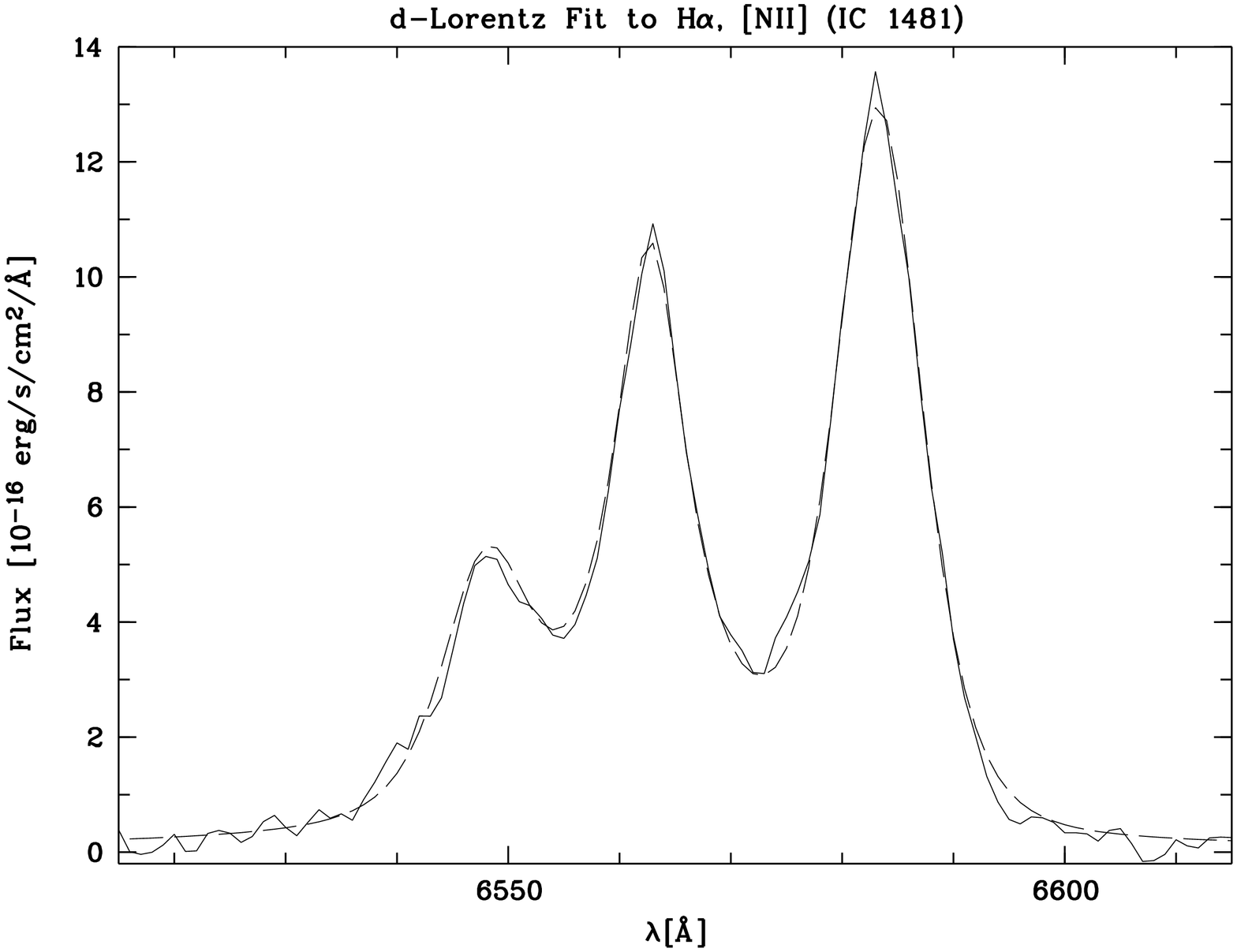}
\caption{\label{ha_n2}
The same as in Fig.~\ref{hb_o3} for the
H$\alpha$ and [\ion{N}{ii}]\,$\lambda\lambda$6548,6583
blend (\object{IC\,1481}). 
Again, $d$--Lorentzians lead to the best total fit.}
\end{figure}

It turns out that all line profiles from the galactic nuclei had stronger wings 
than a single Gaussian and slightly weaker wings than a single Lorentzian, but could 
usually be well fit by a $d$--Lorentzian, which also allowed an appropriate decomposition 
of the H$\alpha$+[NII] blend.  
In Table~\ref{fwhm}, heliocentric velocities and line widths (FWHM) as
well as mean $d$--parameters as
derived from $d$--Lorentz fits are shown for the six strongest lines
(excluding [\ion{O}{ii}]\,$\lambda$3727, as it is an unresolved double line).
Examples of the fits are shown in Figs.~\ref{hb_o3} and
\ref{ha_n2}.

From the parameters of the $d$--Lorentz fits of strong lines, the line intensities relative 
to H$\beta$ were derived (see Appendix~\ref{app}) while the intensities of
faint (single) lines were 
directly obtained by integration using the MIDAS command ``integrate/line''. As for \object{IC\,1481},
extinction was derived with the $c$--method \citep{ost89} by using the recombination value 
for the intensity ratio H$\alpha$/H$\beta$ = 3.1 and an average reddening curve. Both 
observed and reddening--corrected line--intensity ratios from the central
region ($2\arcsec \times 
2\arcsec$) are presented in Table~\ref{ratio}. For pairs of lines ([\ion{O}{iii}], 
[\ion{O}{i}], and [\ion{N}{ii}]) with a fixed line ratio ($\sim$\,3:1), only the brighter 
line is listed. Estimated errors of the line ratios lie in the range of 10\% $-$ 20\%.
Comparison with literature was only possible for the brighter lines of \object{ESO\,103--G035} 
\citep{mor88}, as all other line ratios have not been reported in the literature yet. Deviations 
of observed line ratios lie in the range of $\sim$\,1\,\% $-$ 35\,\% 
and may be due to the fact 
that \citet{mor88} did not use a template spectrum to subtract the 
underlying stellar contribution. Furthermore, they
refer to a larger region (1\farcs5 $\times$ 5\arcsec). 

Applying the classical methods outlined in \citet{ost89}, we derived the electron 
density using [\ion{S}{ii}]\,$\lambda$6716/$\lambda$6731 in the central region
($2\arcsec \times 2\arcsec$). The temperature, determined from [\ion{O}{iii}] 
$\frac{\lambda 4959+\lambda 5007}{\lambda 4363}$, only corresponds to the brightest 
row (formally $0\farcs68 \times 2\arcsec$) because of the weakness of the $\lambda4363$ 
line.  Clearly, these parameters given in Table~\ref{result} represent some average over
the central several hundred parsecs in each case (see Table~\ref{sample} for the typical 
scale in each galaxy). For \object{TXS\,2226--184} and \object{IC\,1481}, no temperature was measured, 
since the [\ion{O}{iii}]\,$\lambda$4363 emission line was too faint. In the spectrum of 
\object{IC\,1481}, the [\ion{S}{ii}]\,$\lambda$6731--emission line was truncated by telluric 
absorption bands, thus no electron density could be derived. For \object{TXS\,2226--184}, 
[\ion{S}{ii}]\,$\lambda$6716 might be slighly affected by the telluric absorption bands 
and the deduced electron density has to be taken with some caution. The reddening--corrected
H$\beta$ luminosity for the central region ($2\arcsec \times 2\arcsec$) 
is given in Table~\ref{result} as well.

The most important line ratios from the central region 
($2\arcsec \times 2\arcsec$) which distinguish 
between emission--line object classes were calculated (see Table~\ref{result}) and put in the 
diagnostic diagrams as used in \citet{ost89} and \citet{bal81}. \object{ESO\,103--G035} falls into the 
Seyfert regime. As the spectra of \object{ESO\,103--G035} did not show any broad components neither in 
H$\alpha$ nor in H$\beta$ in agreement with the results of \citet{mor88}, we classify \object{ESO\,103--G035} 
as Seyfert--2 galaxy.

Line ratios of \object{TXS\,2226--184} and \object{IC\,1481} follow closely the LINER classification of \citet{hec80}
([\ion{O}{iii}]/[\ion{O}{ii}] $\leq$ 1 and [\ion{O}{iii}]/[\ion{O}{i}] 
$\leq$ 3) and fall into the LINER regime in both the line--ratio diagrams used by \citet{ost89} 
and \citet{bal81}. Furthermore, this classification is in agreement with their
low luminosity, and, for \object{IC\,1481}, identical with the results from \citet{huc82}.
Our classification of \object{TXS\,2226--184} as a LINER is new.
\section{Discussion}
\subsection{Why $d$--Lorentzians?}
When discussing emission--line profiles of active galaxies, it is important
to keep in mind, that they might represent line--of--sight integrations
of several kinematical components such as rotating disks, cones
of outflowing gas or expanding shells.
\citet{sch95} have shown that line--of--sight integrations 
through such configurations would neither yield Gaussians nor Lorentzians.
Furthermore, if the spatial and spectral smearing by an instrumental function
is wide compared to the kinematical gradients, the profiles will reveal
a more symmetric core--wing structure \citep{sch95}.
The latter also applies here with
our spectral resolution of $\sim$\,130\,km\,s$^{-1}$ and spatial resolution of $\sim$\,1\farcs2
- 2\farcs5 which lead to an integration over relatively large cells of the phase
space of an NLR.

Various observations revealed that even the ``narrow'' lines in AGNs
can show considerable profile structure
if measured at sufficient resolution \citep{vrt85,whi85,sch03}.
Commonly, in the case where single--component fits fail,
multi--component Gauss fits or Lorentz fits are employed.

While Gaussians are relatively well suited to describe instrumental
functions, thermal Doppler broadening or some kind of turbulence,
they often fail to fit broad wings. They can instead
be well fit by Lorentzians, which differ from Gaussians with same FWHM
by more extended wings.
\citet{whi85} already described the non--Gaussian nature of
observed [\ion{O}{iii}] line profiles which revealed ``a stronger base 
relative to the core than Gaussians''.
\citet{ver01} also reported that most broad emission lines of 
Narrow--Line Seyfert 1 (NLS1) galaxies could be well fit by a single Lorentz
profile, confirming previous claims that Lorentzians rather than Gauss profiles are 
better suited to reproduce the shape of the NLS1's broad emission lines. \citet{sch03} 
showed that Lorentzians appear to be better suited to fit the spectra of NGC\,1052 
and Mrk\,1210. Hence, extended wings of intrinsic narrow--line profiles
of the bulk components may not be uncommon.

However, the use of $d$--Lorentzians is new and allows us to fit 
both permitted and forbidden lines by adjusting the additional parameter $d$ 
(see Appendix~\ref{app} for consequences concerning the FWHM).
Note that, as we are dealing here with Seyfert--2s and LINERs, no broad lines
were fit. In those cases in which 
both broad-- and narrow--line components can be
clearly distinguished, a multi--component fit
is of course justified.
According to Occam's razor, the use of $d$--Lorentzians 
has the advantage of adding just a single parameter instead of using
multi--component fits of Lorentzians and Gaussians.
As $d$--Lorentzians seem to provide a very good fit to the data, the emission--line fluxes
can be well approximated as the integral over the corresponding $d$--Lorentzian,
which is a great advantage especially when measuring blended lines. 
Using $d$--Lorentzians does not imply a loss 
of physical information: On the one hand,
both Lorentzians and Gaussians might not
be physically motivated in all cases
(as argued above) and on the other hand,
the parameter $d$ gives a measure of the width of the line wings.

Our observed line wings have typically reached the noise
levels at $\sim$\,6\% of the maximum.
Using equations (A.1) $-$ (A.3), a Gaussian falls to 6.3\% of the maximum
at 2 HWHMs (half width at half maximum) from the center, a Lorentzian to 5.9\%
at 4 HWHMs and a $d$--Lorentzian has decreased to 5.9\% at 2.8 HWHM, taking a typical
value of $d$ = 1.3. Using a mean FWHM 
from the six strongest lines of the three galaxies (Table~\ref{fwhm}),
the line wings have velocities of 600 $-$ 700\,km\,s$^{-1}$.
Thus, the question arises how such large velocities
in the line wings can be attained.
This was discussed by \citet{sch03} 
in terms of the presence of turbulence, outflows, 
magneto--hydrodynamic waves and electron scattering.
They concluded that the latter might provide a viable explanation
of the observed line wings.

For our new sample of galaxies, we may ask again, 
whether electron scattering can produce the observed 
line wing velocities, either due
to the presence of hot coronal or cooler ionized gas.
The ratio of scattered to input luminosity
can be estimated as $L_s$/$L_{in}$ = $f \tau_e$,
with $\tau_e$ = electron--scattering optical depth and
$f$ = covering fraction of the scattering medium.
To obtain $\tau_e$ on the percent level,
effective column--densities of fully ionized gas of $2 \cdot
10^{22}$\,cm$^{-2}$ would be required.
The total bremsstrahlung luminosity of
a kpc--extended sphere of a hot intercloud medium with such a column density
can be estimated as
$2.6 \cdot 10^{42} (n_e / 10 {\rm cm}^{-2})^2 \sqrt{T/10^6 {\rm K}}(R/500 {\rm
pc})^3$\,erg\,s$^{-1}$ (Eq. 5.15b in \citet{ryb79}).
The intrinsic (unabsorbed) thermal soft--X--ray emission
for \object{ESO\,103--G035} (the only galaxy in our sample with measured 
0.5--2 keV flux) is estimated to be $\sim$\,4
$\cdot$ 10$^{42}$\,erg\,s$^{-1}$ \citep{tur97}.
For a temperature of $T \sim 10^6$, a density of $n_e \sim 10^2$
and a radius of $R \sim 1$\,kpc, a luminosity $\sim 500$ times higher
would be expected.
Thus, as in \citet{sch03} for NGC\,1052 and Mrk\,1210, 
we do not see enough hot coronal gas to produce the required
wings in \object{ESO\,103--G035}. 
However, electron scattering inside the cooler ionized gas component
($T \sim 10^4$\,K) with electron densities of $n_e \sim 500$
(which were observed for \object{ESO\,103--G035}, Table~\ref{result})
is a possible explanation.
Calculating the expected soft--X--ray luminosity
with a scale of $R \sim 10^2$\,pc yields $\sim 5 \cdot$
10$^{42}$\,erg\,s$^{-1}$, in agreement with the measurements
of \citet{tur97}.
Considering a geometrical
scattering factor $f \sim 0.3$, one easily obtains 
$L_s$/$L_{in}$ of a few percent, which suffices to explain the extended wings
of the narrow lines. Thus, electron scattering even at the cooler
ionized gas can account for the observed velocities in the line wings.
\subsection{What triggers the water megamaser activity?}
The unique association of H$_2$O megamasers with AGNs of Seyfert--2 and LINER
type and the fact that the emission originates from the innermost parsec(s)
of the parent galaxy \citep{gre95a,miy95,cla98,tro98,her99}
suggest that the unknown excitation mechanism is closely
related to the AGN activity 
(and that at least some LINERs are AGNs rather
than starbursts).
The Seyfert--2 geometry is favorable for megamaser activity supplying
the necessary large column densities of warm and dense molecular gas enriched
with H$_2$O molecules. 

Not in all cases, however, a nuclear radio continuum background 
has to be amplified to obtain detectable megamaser emission 
(e.g. \citet{gre95a,gre95b}). Sometimes, column densities of a disk
may be large enough even when viewed face--on and little is known 
about the solid angle of the emitted maser radiation. It is thus 
possible that in exceptional cases nuclear maser emission is also 
detectable in Seyfert--1 galaxies \citep{nag02,hag03}.

Interaction of molecular gas with radio jets will be associated
with shocks and the shock heated gas favors H$_2$O maser emission
(e.g. \citet{pec03}). Optically detectable ionized outflowing gas 
may, however, also trigger maser emission. 
\citet{fal01} reported linear jet--like HST features in 
\object{TXS\,2226--184} and NGC\,1386 that might
be related to an outflow rather than an excitation cone.
In the megamaser sample of 
\citet{sch03}, galactic rotation and outflow of narrow--line 
gas are common features. The detection of H$_2$O megamasers in very 
luminous infrared galaxies ($L_{\rm FIR}$ $>$ 10$^{11}$\,L$_{\odot}$; 
\citet{hag02a,pec04}) with
presumably many young clusters of O and B stars may add further support 
for such an outflow scenario.
\subsubsection{Three specific cases}
There are a few nearby H$_2$O megamaser sources that are particularly 
well studied, not only in view of their H$_2$O maser properties but
also at other wavelengths, including the optical and near infrared 
windows. Best known are the three northern of the five originally 
discovered megamaser sources, NGC\,1068, NGC\,4258 
\citep{cla84} and NGC\,3079 \citep{hen84,has85}. 
The H$_2$O data from NGC\,4258 show a warped 
nuclear accretion disk that is seen almost edge--on (e.g. 
\citet{miy95}). Towards NGC\,1068, we may view a torus with ``sub--Keplerian'' rotation
curve \citep{hur02} that might be more massive, also
with respect to the nuclear engine, and that appears to be thicker or slightly less 
well ordered. To the north, where the nuclear jet is bending, jet--masers
are observed 
(e.g. \citet{gal01}). NGC\,3079
appears to show a disk that is less well ordered than those of NGC\,1068 
and NGC\,4258, but a detailed map of the red--shifted maser features is 
still needed to obtain a complete picture \citep{tro98,hag02b}.

Direct hints for the presence of megamaser emission in these sources
is given by the near infrared continuum, providing evidence for the 
presence of warm dust heated by the AGN. NGC\,3079 harbors a compact 
(1\arcsec) nuclear source with dust at a temperature of $\sim$\,1000\,K 
\citep{isr98}. An optically detected large scale outflowing
zone \citep{cec01} is apparently not triggering 
maser emission. An even more compact dusty core is detected in NGC\,4258,
with an upper size limit of 200\,mas (7\,pc; \citet{cha00}). 
Two jets (anomalous arms) reach out of the nuclear region and 
shock the ambient gas in the inner 350\,pc 
\citep{wil01}. Interestingly, the jets are much more prominent than in most other 
spirals, but do not trigger any megamaser emission. Instead, they provide 
the radio continuum background for the enhanced flux of the systemic 
H$_2$O features that are stronger than the presumably self--amplified 
red-- and blue--shifted components 
\citep{her98}. The best studied galaxy is NGC\,1068 
(e.g. \citet{gal03}). This prototypical Seyfert--2 galaxy also contains a 
compact near-- and mid--infrared source 
\citep{mar97,rou98}. The extremely red colors of its 200\,mas 
core (15\,pc) lead to an intrinsic extinction of $A_{\rm V}$ $\ga$ 
25$^{\rm m}$, assuming classical grains at 1500\,K. Optical 
spectroscopy with the HST shows lines
split into two velocity components separated by 1500\,km\,s$^{-1}$
within the inner arcsec \citep{axo98}. 
2 micron H$_2$ emission is also double peaked \citep{gal02}. 
Both can be explained by gas interacting with the radio jet.

To summarize, the three galaxies NGC\,1068, NGC\,4258, and
NGC\,3079 exhibit a spatially compact near 
infrared core containing dust clouds that are heated by the central 
engine. This appears to be the main hint for the potential presence of
accretion disk masers, but these cores may be more difficult to detect 
than the masers. 
Tracers for jet masers are split lines at optical
and near infrared wavelengths while outflow masers should show broad wings
in lines arising from ionized gas as e.g. seen in NGC\,1052 and Mrk\,1210 
\citep{sch03}. The nature of these outflows has, however, to 
be clearly identified to distinguish them from gas entrained by the much 
faster nuclear jets.
\subsubsection{Our data}
The new data presented here support the connection between H$_2$O
megamaser emission and the presence of Seyfert--2 and LINER nuclei.
They do not, however, further elucidate the relationship between
outflows and megamasers as they do not show signs of outflow
in their optical spectra. In any case, establishing a connection 
between optical and radio data is difficult keeping in mind the 
different angular scales involved (i.e. a few milliarcseconds
for 1.3\,cm interferometric maps and a few arcseconds for our study), so that
optical and near infrared data taken with subarcsec resolution would be 
highly desirable.
\subsubsection{H$_2$O maser classification}
We conclude that all 
22\,GHz H$_2$O masers detected so far might be subdivided into five categories 
with rising but often overlapping ranges of luminosity: 
(i) Masers from late--type stars, (ii) masers from star forming regions,
(iii) nuclear ``outflow'' masers in which outflows impinging onto dense
molecular clouds may provide a suitable trigger for kilo-- or 
megamaser emission (possibly occurring in starburst galaxies,
LINERs dominated by starburst instead of AGN activity, and type--1 AGNs),
(iv) ``jet'' masers where a direct interaction 
between the nuclear jet and the interstellar medium triggers maser activity and (v) 
``accretion disk'' masers where the masers are aligned in sometimes 
warped disks or thick tori.
\section{Conclusions}
We analyzed optical spectra of the megamaser galaxies
\object{ESO\,103--G035}, \object{TXS\,2226--184}, and \object{IC\,1481} and 
presented rotation curves, line ratios,  
electron densities, temperatures, and $L_{\rm H\beta}$.
The successful line--profile decompositions 
of single lines as well as blends
rest on $d$--Lorentzians
with an additional parameter $d$ to adjust the wings, rather than Gaussians
or Lorentzians as basic functions.
Using $d$--Lorentzians, we can fit the extended lines
of intrinsically narrow--line profiles. Crude estimates suggest that
electron scattering at the ionized gas itself might lead
to a viable explanation of the large line widths
observed, but detailed modelling of such
processes would be useful.
On the basis of line ratios and their low luminosity,
\object{TXS\,2226--184} and \object{IC\,1481} are classified 
as LINERs while \object{ESO\,103--G035}
is a Seyfert 2. No significant velocity gradient is observed along the major axis
within the inner 2\,kpc of \object{TXS\,2226--184}.
\object{IC\,1481} reveals a post--starburst spectrum
which could not be fit by an S0 template galaxy only.
Instead, using additional A0V $-$ F8V star spectra, we could
correct for the Balmer absorption lines.

The comparison of optical spectra of all three H$_2$O megamaser 
galaxies with an angular resolution of $\sim$\,2\arcsec~support
the connection between H$_2$O megamaser emission 
and the presence of Seyfert--2 and LINER nuclei. However, the spectra do not provide 
direct clues to the very nuclear regions where the masers reside.
Their classification as
``outflow'', ``jet'' or ``accretion disk'' masers remain uncertain.
Apparently, either the nuclei are too heavily
obscured by dust associated with cool dense foreground gas or the
morphological structures giving rise to the megamaser emission, i.e.
accretion disks and jets, have linear scales that are far too small 
to be detectable with an arcsec resolution. 

\begin{acknowledgements}
In memoriam Prof. Hartmut Schulz, deceased in August 2003.
N.B. remembers him gratefully for having been her 
``Doktorvater'' in the truest sense of the word.
C.H. wants to thank his friend
and co--author for a fruitful and enjoyable collaboration over 
almost 10 years.
The astronomical community has lost a 
highly honourable and agreeable colleague.
N.B. is grateful for financial support of the ``Studienstiftung
des deutschen Volkes''. We would like to thank the anonymous referee
for useful comments and suggestions.
This research has made use of the NASA/IPAC Extragalactic Database
(NED), which is operated by the Jet Propulsion Laboratory,
Caltech, under contract with the 
NASA.
\end{acknowledgements}

\appendix
\section{Line--profile fits}
\label{app}
\begin{figure}[h!]
\centering
\includegraphics[width=6cm,angle=-90]{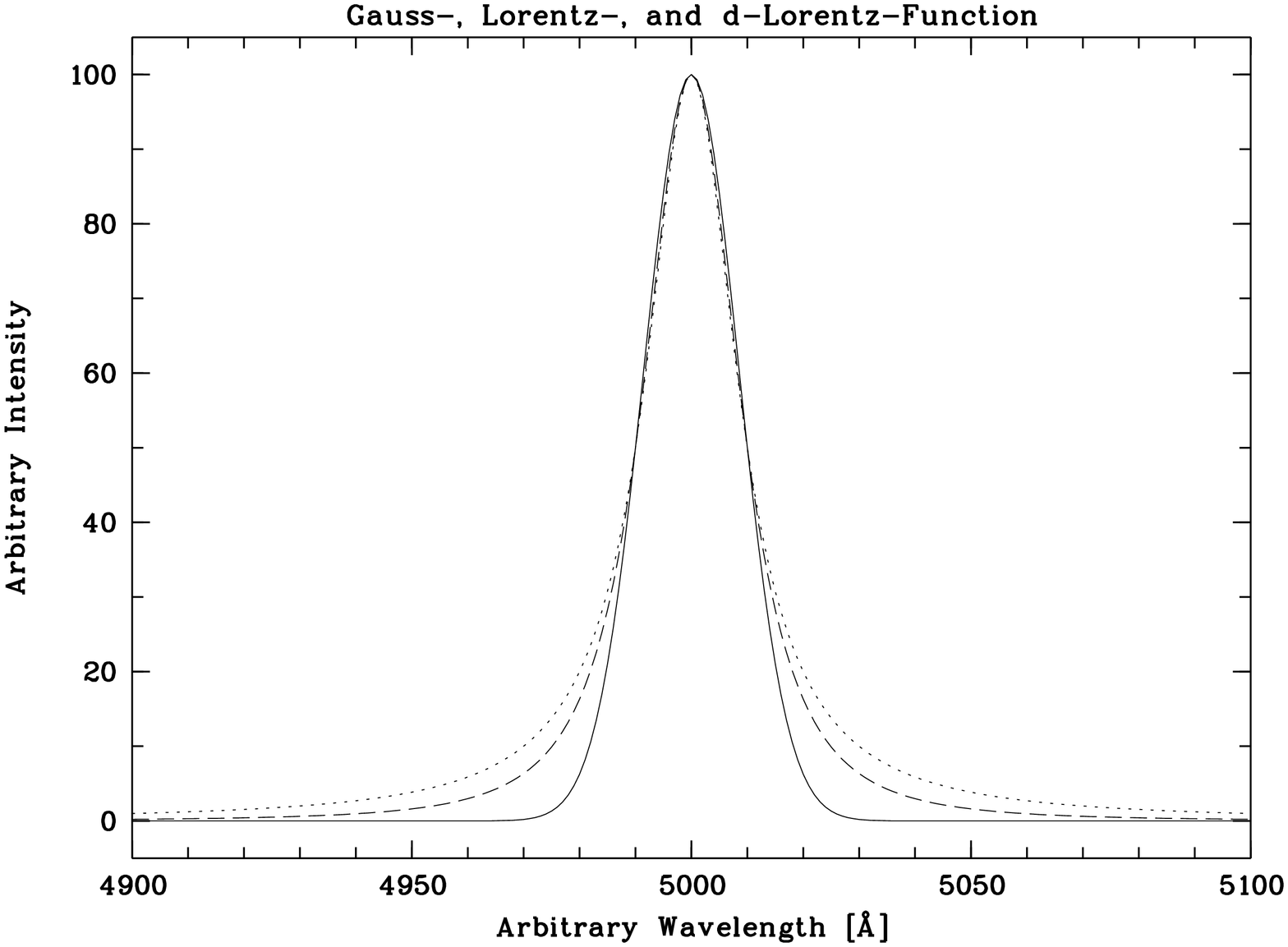}
\includegraphics[width=6cm,angle=-90]{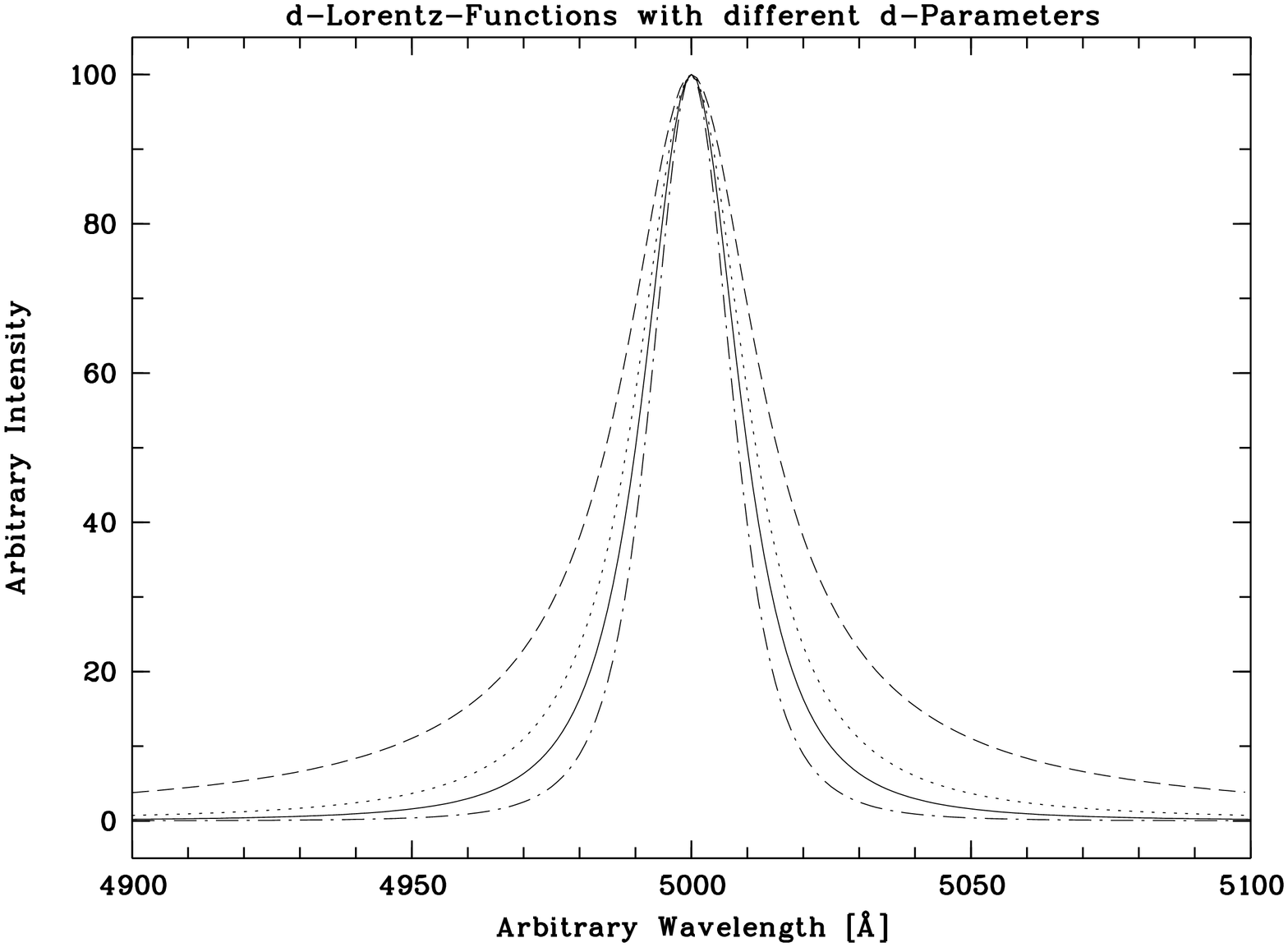}
\caption{\label{vgl} Upper panel:
Comparison of Gaussian (solid), Lorentzian (dotted) and $d$--Lorentz functions (dashed 
line) ($a$ = 100, $b$ = 5000, $c$ = 20 for Gaussian and Lorentzian, $c$ = 26.1 
and $d$ = 1.5 for $d$--Lorentzian to gain the same effective FWHM). A $d$--Lorentzian 
falls in between a Gaussian and a Lorentz function. Lower panel: Four $d$--Lorentzians 
are shown with different $d$ parameters [$d$ = 0.8 (dashed), $d$ = 1.2 (dotted), 
$d$ = 1.5 (solid) and $d$ = 2 (dash--dotted)]. With increasing $d$, the wing profile 
narrows.  Thus, the parameter $d$ can be used to adjust the wings.}
\end{figure}

Line profiles of the strongest lines have been fit with Gaussians, Lorentzians,
and $d$--Lorentz functions as provided by MIDAS using the same initial parameters
and the commands ``edi/fit'' and ``fit/image''. These basic fit functions are 
compared in Fig.~\ref{vgl}, where it is shown that Lorentzians have much stronger
wings than Gaussians, but that the wings of $d$--Lorentzians can be adjusted 
to intermediate wing strengths. The functions are defined as ($a$ = height, 
$b$ = position, $c$ = FWHM for Gaussian and Lorentzian, and $d$ = additional 
parameter for a $d$--Lorentzian)
\begin{eqnarray}
{{\rm Gauss}}\, (x;a,b,c) = a \exp \left[- \ln 2 
\left(\frac{2(x-b)}{c}\right)^2\right]\label{gau}\\ 
{{\rm Lorentz}}\, (x;a,b,c) = a \left[1 + 
\left(\frac{2(x-b)}{c}\right)^2\right]^{-1}\label{cau}\\
{d\rm{-Lorentz}}\, (x;a,b,c,d) = a \left[1 + 
\left(\frac{2(x-b)}{c}\right)^2\right]^{-d}\label{lor} \hspace*{0.0cm} .
\end{eqnarray}
\noindent
In all galaxies, all spectral rows and for all analyzed strong emission--lines,
a Gaussian yields too narrow wings, a Lorentzian too broad ones and a $d$--Lorentzian 
leads to the best fit (for comparison, two examples are plotted in
Figs.~\ref{hb_o3} 
and~\ref{ha_n2}).  In a $d$--Lorentzian, $c$ is not the FWHM width, which is instead given 
as a combination of $c$ and $d$,
\begin{eqnarray}\label{b50}
{\rm FWHM}_{d{\rm-Lorentz}} = c \cdot \sqrt{2^{1/d} - 1} \hspace*{0.2cm} .
\end{eqnarray}
The width at $n$ \% of the height can be calculated with
\begin{eqnarray}\label{b20}
{\rm W}n_{d{\rm-Lorentz}} = c \cdot \sqrt{\left(\frac{1}{n}\right)^{1/d} - 1}
\hspace*{0.2cm} .
\end{eqnarray}
With increasing $d$ the wing--profile narrows (see Fig.~\ref{vgl}). The 
line intensities can be derived by computing the following integral
\begin{eqnarray}
I_{d{\rm-Lorentz}} &=& a \cdot c \cdot \frac{\sqrt{\pi}}{2} \cdot \frac{\Gamma 
(-0.5 + d)}{\Gamma (d)}\label{ilor} \hspace*{0.2cm} .
\end{eqnarray}

\end{document}